\documentclass[a4paper,11pt]{article}
\pdfoutput=1 
\usepackage{subcaption}

\usepackage{jinstpub} 
\usepackage{lineno}
\def\myhyphen{{\hbox{-}}}
\usepackage{booktabs}
\usepackage{amsmath}
\title{\boldmath Measurement of cosmic-ray reconstruction efficiencies in the MicroBooNE LArTPC using a small external cosmic-ray counter}

\collaboration{MicroBooNE Collaboration}
\author[g]{R.~Acciarri}
\author[cc,h]{C.~Adams}
\author[i]{R.~An}
\author[c]{J.~Anthony}
\author[z]{J.~Asaadi}
\author[a]{M.~Auger}
\author[g]{L.~Bagby}
\author[cc]{S.~Balasubramanian}
\author[g]{B.~Baller}
\author[o]{C.~Barnes}
% ...

\affiliation[a]{Universit{\"a}t Bern, Bern CH-3012, Switzerland}
% ...
\affiliation[c]{University of Cambridge, Cambridge CB3 0HE, United Kingdom}
% ...
\affiliation[g]{Fermi National Accelerator Laboratory (FNAL), Batavia, IL 60510, USA}
\affiliation[h]{Harvard University, Cambridge, MA 02138, USA}
\affiliation[i]{Illinois Institute of Technology (IIT), Chicago, IL 60616, USA}
% ...
\affiliation[o]{University of Michigan, Ann Arbor, MI, 48109, USA}
% ...
\affiliation[z]{University of Texas, Arlington, TX, 76019, USA}
% ...
\affiliation[cc]{Yale University, New Haven, CT, 06520, USA}

\abstract{The MicroBooNE detector is a liquid argon time projection chamber at Fermilab designed to study short-baseline neutrino oscillations and neutrino-argon interaction cross-section. Due to its location near the surface, a good understanding of cosmic muons as a source of backgrounds is of fundamental importance for the experiment. We present a method of using an external 0.5 m (L) x 0.5 m (W) muon counter stack, installed above the main detector, to determine the cosmic-ray reconstruction efficiency in MicroBooNE.
Data are acquired with this external muon counter stack placed in three different positions, corresponding to cosmic rays intersecting different parts of the detector.
The data reconstruction efficiency of tracks in the detector is found to be $\epsilon_{\mathrm{data}}=(97.1\pm0.1~(\mathrm{stat}) \pm 1.4~(\mathrm{sys}))\%$, in good agreement with the Monte Carlo reconstruction efficiency $\epsilon_{\mathrm{MC}} = (97.4\pm0.1)\%$.
This analysis represents a small-scale demonstration of the method that can be used with future data coming from a recently installed cosmic-ray tagger system, which will be able to tag $\approx80\%$ of the cosmic rays passing through the MicroBooNE detector.}


\begin{document}
\maketitle
\flushbottom

\section{Introduction}
\label{sec:intro}
MicroBooNE (Micro Booster Neutrino Experiment) is a liquid argon time projection chamber (LArTPC) located at the Fermi National Accelerator Laboratory (Fermilab) \cite{detector}. The main physics goals of the experiment are to investigate the excess of low-energy events observed by the MiniBooNE collaboration \cite{miniboone} and to measure neutrino-argon interaction cross sections. MicroBooNE also provides important research and development contributions to detector technology and event reconstruction techniques for future LArTPC experiments, such as DUNE (Deep Underground Neutrino Experiment) \cite{dune}. The MicroBooNE detector is located 470 m downstream of the Booster Neutrino Beam (BNB) target. The BNB is predominantly composed of muon neutrinos ($\nu_{\mu}$) with a peak neutrino energy at about 0.7 GeV.

\begin{figure}[htbp]
  \begin{center}
    \includegraphics[width=0.8\linewidth]{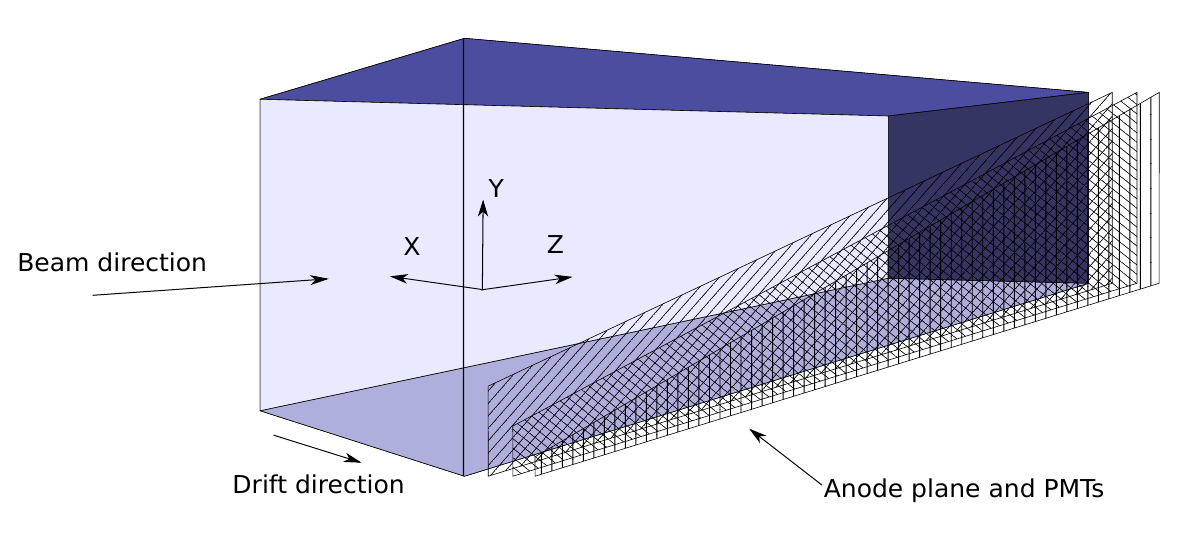}

    \caption{The MicroBooNE coordinate system. The three wire planes, shown in the right front face, are vertical (collection plane) and at  $\pm60^{\circ}$ to the vertical (induction planes). The dimensions of the TPC are 256 cm $\times$ 233 cm $\times$ 1037 cm in the $x$, $y$, and $z$ directions, respectively.} \label{fig:coord}
  \end{center}
\end{figure}

The MicroBooNE detector  consists of a rectangular time projection chamber (TPC) with dimensions of 256 cm (width) $\times$ 233 cm (height) $\times$ 1037 cm (length).  The cylindrical cryostat contains a total of 170 t of liquid argon, while the mass of liquid argon in the active volume, defined as the portion of the argon encompassed by the TPC, is 89 t. Figure~\ref{fig:coord} shows a graphical representation of the TPC in the MicroBooNE coordinate system. The $x$ direction corresponds to the drift coordinate, the $y$ direction is the vertical direction, and the $z$ direction points along the beam.

The TPC consists of three wire planes with 3 mm spacing at angles of $0^{\circ}$,  $+60^{\circ}$ and  $-60^{\circ}$ with respect to the vertical. The cathode, made of a plane of stainless steel panels, operates at a voltage of $-70$ kV. In a neutrino interaction, a neutrino from the beam interacts with an argon nucleus, and the secondary charged particles traverse the medium, losing energy and leaving an ionization trail. The resulting ionization electrons drift to the wire planes under an electric field of 273 V/cm. The distance between the cathode and anode is 2.56 m. An ionization electron takes about 2.3 ms to travel the full drift distance, called the drift time window. Charge drifting past a wire plane induces a current that produces a bipolar signal in the electronics. The first two planes are referred to as induction planes. The wire plane furthest from the cathode has wires oriented vertically. Drifting electrons are collected on this plane producing a unipolar signal. Charge deposited in the TPC generates a signal used to create three distinct two-dimensional views (in terms of wire and time) of the event, which can be combined to reconstruct a three-dimensional image of the interaction.
A set of 32 photomultipliers tubes (PMTs) is placed behind the anode plane to collect the argon scintillation light. Scintillation light provides timing information with few-ns precision, which provides the TPC start time of the event and can be used for background suppression. More details about the MicroBooNE detector can be found in ref. \cite{detector}.

The detector is placed in a pit 6 m below the surface with no overburden. The muon cosmic-ray rate in the MicroBooNE detector is estimated to be 5.5 kHz, which corresponds to $\approx13$ muons per TPC drift time window of 2.3 ms. The abundant flux of cosmic muons is a source of background to neutrino events, and an optimal reconstruction of the cosmic rays in the TPC is therefore crucial.

In order to study the challenges of cosmic-ray background rejection in a surface neutrino experiment, the MicroBooNE detector was equipped with an external muon counter stack (MuCS) at the start of operations in 2015. We use this system to develop and demonstrate muon tagging. It also provides an external set of data to validate simulation and reconstruction.
In the future, the method described in this paper will be applied to the data coming from the cosmic ray tagger (CRT) system \cite{crt}, installed in March 2017. It is able to tag approximately $80\%$ of the cosmic rays traversing the MicroBooNE LArTPC, which is an order of magnitude more than the coverage provided by the MuCS. The better coverage the CRT provides allows to determine efficiencies over the full detector volume, to measure e.g. the cosmic-ray flux in the LArTPC. 


\section{The Muon Counter Stack}\label{sec:proc}
\begin{figure}[htbp]
  \begin{center}
    \includegraphics[width=0.7\linewidth]{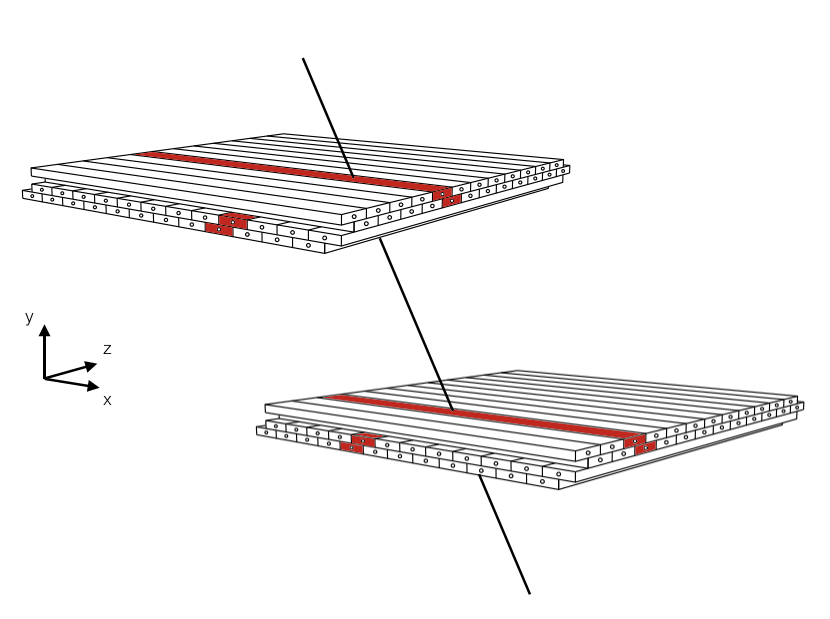}
    \caption{A cosmic ray passing through the MuCS boxes hits the scintillator strips. We have a MuCS hit when the signal corresponding to the strip is above a certain threshold. The position of the panels along the $y$ axis and the position of the hit strips (highlighted in red) are used to extrapolate the trajectory of the cosmic ray down to the TPC located below the counters.} \label{fig:boxes}
  \end{center}
\end{figure}

The Muon Counter Stack, described in detail in ref. \cite{mucs}, consists of two sets of planar modules placed into two separate, light-tight boxes. The upper and lower boxes are placed $2.75$~m and $2.03$~m above the TPC, respectively. Their position is known to a precision of $0.5$~cm. Each planar module is constructed using 48 scintillator strips of $4$~cm width, $48$~cm length, and $2$~cm thickness. The scintillator strips are arranged into a pair of bi-layers, each 12 strips wide and oriented perpendicular to each other. The overall setup is shown in figure~\ref{fig:boxes}.

Each strip contains a wavelength shifting optical fiber, connected to a multi-anode PMT, which is read out by a dedicated DAQ system that records the hit patterns of the scintillator strips.
The MuCS is designed to provide a trigger on through-going muons that intersect all four bi-layers of scintillator strips.

\begin{figure}[htbp]
  \begin{subfigure}{0.30\textwidth}
    \includegraphics[width=\linewidth]{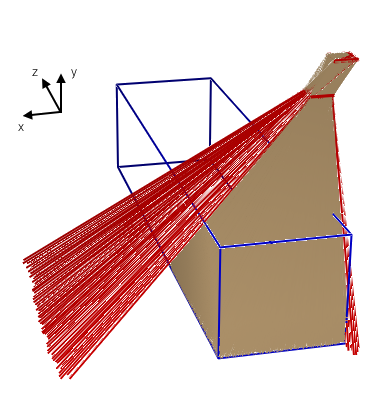}
    \caption{Upstream} \label{fig:upstream}
  \end{subfigure}
  \begin{subfigure}{0.30\textwidth}
    \includegraphics[width=\linewidth]{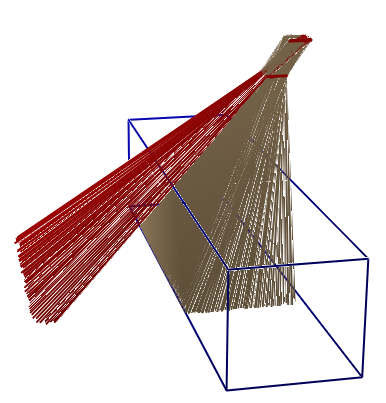}
    \caption{Center} \label{fig:centre}
  \end{subfigure}
  \begin{subfigure}{0.30\textwidth}
    \includegraphics[width=\linewidth]{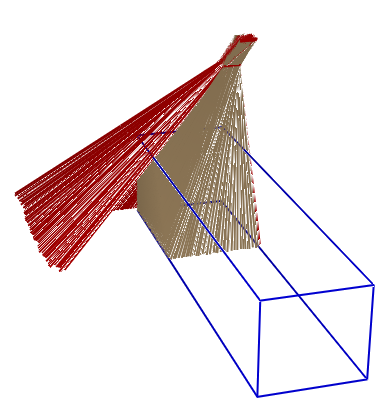}
    \caption{Downstream} \label{fig:downstream}
  \end{subfigure}

  \caption{Illustration of the three MuCS configurations with a Monte Carlo simulation of the possible MuCS trajectories. Brown tracks correspond to cosmic rays hitting both the MuCS boxes and TPC, while red tracks traverse only the MuCS and miss the TPC.} \label{fig:mucs}
\end{figure}

The data used in our analysis has been acquired with the MuCS in three different geometrical configurations. The three configurations correspond to a setup with the two boxes placed above the TPC and at the upstream end, at the center, and at the downstream end of the MicroBooNE detector, keeping the box spacing and alignment identical.
A three-dimensional schematic of the three MuCS setups is shown in figure~\ref{fig:mucs}.

\section{Data reduction and Monte Carlo simulation}\label{sec:merging}

The MuCS trigger is propagated to the MicroBooNE trigger board and provides the starting time ($t_0$) of a track in the TPC associated with the MuCS.
The MuCS triggers at a rate of nearly 3 Hz.
A software filter removes events with more than 4 hit strips per bi-layer from the data sample, discarding events with unclear hit patterns caused by electronic noise or air showers. After this filter, the MuCS data sample consists of $\approx$30000 MuCS-triggered events, acquired in $\sim$10 hours of data taking.
The data follow a processing path that merges the MuCS hit patterns and extrapolated trajectory information with the TPC to form a MuCS-merged dataset. 


As illustrated in figure~\ref{fig:boxes}, when a signal in one strip is above a certain threshold we have a MuCS hit. By combining the MuCS hits in each bi-layer, we obtain two sets of position coordinates of the crossing points of the cosmic rays (the $z$ and $x$ coordinates in the MicroBooNE TPC reference frame shown in figure~\ref{fig:coord}). The height at which the modules are positioned (corresponding to the $y$ coordinate for the MicroBooNE TPC reference frame) allows the extrapolation of a three-dimensional trajectory of the cosmic ray from the MuCS down to the TPC, which is defined as a \emph{MuCS-extrapolated track}.

The starting angle of the cosmic ray trajectory, in spherical coordinates, is defined by:
\begin{align}\label{eq:angles_mucs}
  \theta = \mathrm{acos}\left(\frac{z_{\mathrm{bottom}}-z_{\mathrm{top}}}{r}\right), \quad
  \phi = \mathrm{atan}\left(\frac{y_{\mathrm{bottom}}-y_{\mathrm{top}}}{x_{\mathrm{bottom}}-x_{\mathrm{top}}}\right),
\end{align}
where $r$ is the distance between $(x_{\mathrm{top}},y_{\mathrm{top}},z_{\mathrm{top}})$ and $(x_{\mathrm{bottom}},y_{\mathrm{bottom}},z_{\mathrm{bottom}})$, given by the hit positions in the top and bottom MuCS box, respectively.

In the TPC, ionization electrons from cosmic ray muons passing through the MicroBooNE cryostat are drifted to the wires and TPC hits are extracted, which are then used by the track reconstruction algorithms provided by the Pandora framework \cite{pandora} to form TPC reconstructed tracks.

At the top and bottom boundary of the TPC, this distortion leads primarily to a vertical displacement of the ionization tracks.  The vertical displacements are larger far from the anode (due to the longer travel distance of the ionization electrons) and for positions far from the center of the TPC (where the built-up charge increases the distortion).  The vertical displacement at the top and bottom boundary of the TPC has been measured in the data by reconstructing the start and end points of minimum-ionizing particles crossing the TPC, while the vertical displacement in the TPC bulk has been estimated by a dedicated simulation, described in ref. \cite{spacecharge}. Knowing in this way the magnitude of the effect, it has been possible to correct it in the data, allowing direct comparison with the current Monte Carlo simulation.

The first intersection point between the MuCS-extrapolated track and the TPC is defined as $(x_{\mathrm{MuCS}}$, $y_{\mathrm{MuCS}}$, $z_{\mathrm{MuCS}})$. However, because of multiple Coulomb scattering in the material between the MuCS boxes and the TPC, the starting point of the reconstructed track in the TPC corresponding to the MuCS-triggering cosmic ray does not coincide exactly with the extrapolated intersection point.

The TPC reconstructed track with the closest starting point to the MuCS-extrapolated track intersection point is selected for further analysis and is defined as a \emph{MuCS-tagged track}.
The distance $d$ between the extrapolated intersection point and the starting point of the MuCS-tagged track is defined as
\begin{equation}\label{eq:d}
d = \sqrt{(x_{\mathrm{MuCS}}-x_{\mathrm{reco}})^2+(y_{\mathrm{MuCS}}-y_{\mathrm{reco}})^2+(z_{\mathrm{MuCS}}-z_{\mathrm{reco}})^2},
\end{equation}
where $(x_{\mathrm{reco}},y_{\mathrm{reco}},z_{\mathrm{reco}})$ is the starting point of the MuCS-tagged track.

\begin{figure}[htbp]
  \begin{center}
  \includegraphics[width=0.50\linewidth]{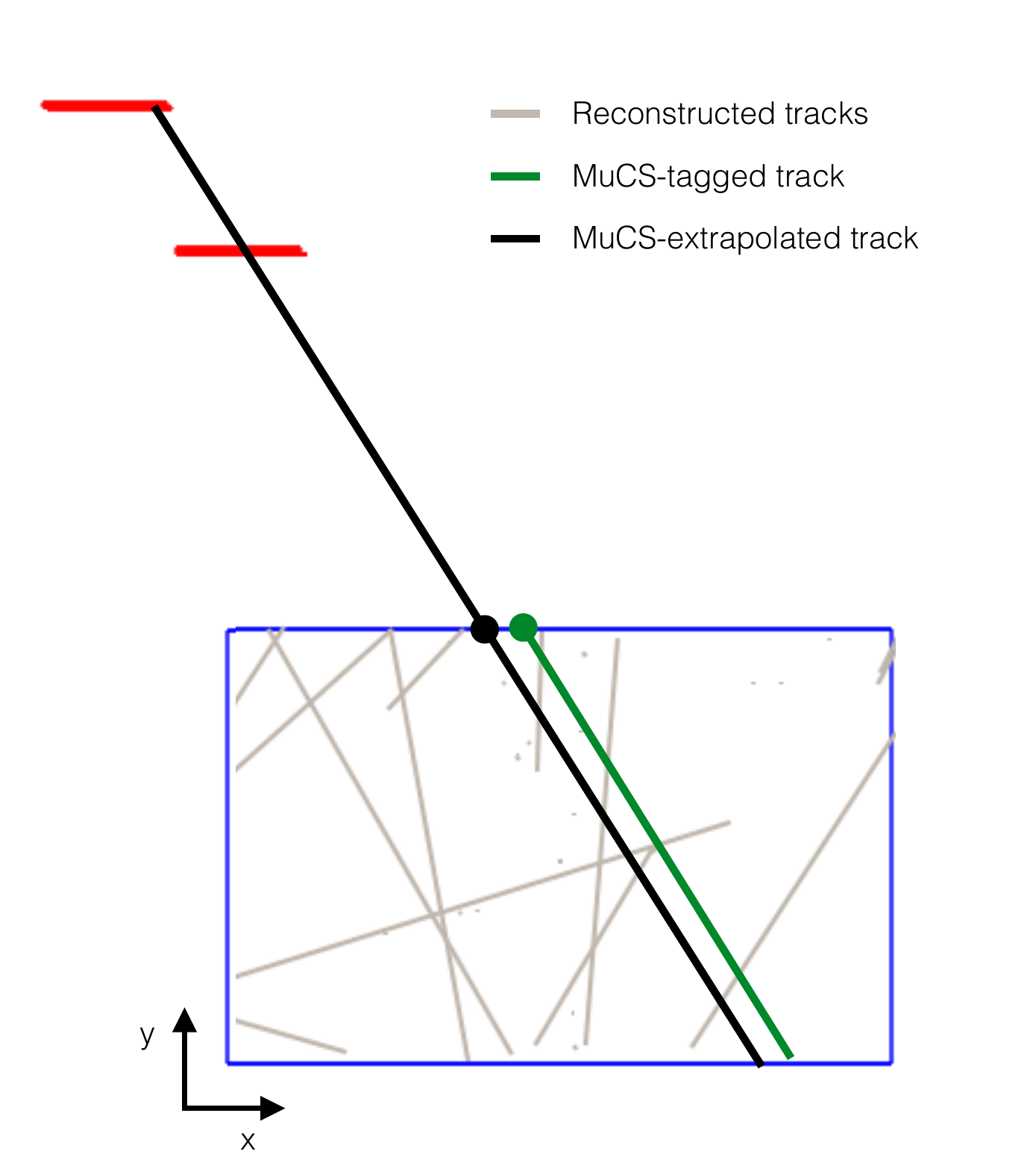}
  \includegraphics[width=0.40\linewidth]{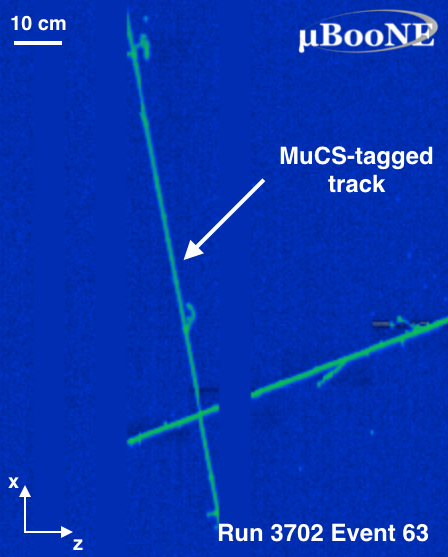}

  \caption{Left: Two-dimensional schematic view of a MuCS event in MicroBooNE. The black line shows the MuCS-extrapolated track, while the green line corresponds to the MuCS-tagged track. The black and green dots correspond to the $(x_{\mathrm{MuCS}},y_{\mathrm{MuCS}},z_{\mathrm{MuCS}})$ and $(x_{\mathrm{reco}},y_{\mathrm{reco}},z_{\mathrm{reco}})$ coordinates, respectively. Right: example of a MicroBooNE event display for the collection plane, showing a MuCS-tagged track in a data event.} \label{fig:evd}
\end{center}
\end{figure}
The data will then include two different sets of information:
\begin{itemize}
  \item MuCS-extrapolated information: a line crossing the entire TPC is extrapolated from the two points given by the MuCS (one for each box). From this extrapolated line it is possible to obtain: (1) the two extrapolated starting angles $\theta$ and $\phi$ of the MuCS, defined in eq. \eqref{eq:angles_mucs}, (2) the extrapolated start point $(x_{\mathrm{MuCS}},y_{\mathrm{MuCS}},z_{\mathrm{MuCS}})$ described above, and (3) the extrapolated end point, corresponding to point where the MuCS-extrapolated track exits the TPC. The extrapolated track length $L$ is calculated by measuring the distance between the extrapolated start point and the extrapolated end point.
  \item Reconstructed TPC data information: for each MuCS-triggered event, the reconstructed starting point $(x_{\mathrm{reco}},y_{\mathrm{reco}},z_{\mathrm{reco}})$ of the MuCS-tagged track.
\end{itemize}
To remove events where the cosmic-ray muon triggered the MuCS but did not hit the TPC, or crossed it for a very short path, we require an extrapolated length in the TPC of $L > 20$ cm.

\subsection{Monte Carlo sample generation}\label{sec:mcgen}

The Monte Carlo sample consists of a simulation of cosmic ray events in the MicroBooNE TPC. The cosmic rays are generated using the CORSIKA \cite{corsika} simulation software. The muons are propagated through the detector using GEANT4 \cite{geant} and passed through a detector simulation stage. The detector simulation reproduces the electron drift, the induction and collection of signals on wires, and the electronics response. The simulation also includes information on the state of the detector readout. Noisy or unresponsive wires, for example, can complicate track reconstruction. The impact of their effect is discussed in section \ref{sec:wires}.


The direction of the simulated cosmic ray is given by its momentum when it enters the TPC, given by GEANT4. The starting angles $\theta$ and $\phi$ are defined in this case as

\begin{align}\label{eq:angles_mc}
  \theta = \mathrm{acos}\left(\frac{p_{z}}{p}\right), \quad
  \phi = \mathrm{atan}\left(\frac{p_{y}}{p_{x}}\right),
\end{align}
where $p_{x}$, $p_{y}$, $p_{z}$ are the $x$, $y$, $z$ components of the cosmic-ray momentum of magnitude $p$.
The track length $L$ is calculated by extrapolating a straight line through the TPC in the $\theta, \phi$ direction and measuring the distance from the true entering point to the extrapolated exiting point in the TPC.

This Monte Carlo simulation provides cosmic rays entering the TPC from all possible directions, while the cosmic rays triggering the MuCS can have only $\theta$, $\phi$ starting angles within the geometrical constraints of the system. Thus, cosmic rays in the Monte Carlo dataset are selected to match the ($\theta$, $\phi$) parameter space covered by the MuCS-extrapolated tracks in data and the Monte Carlo events have been weighted to match the data distributions. 


\section{Reconstruction efficiencies}\label{sec:reco}

The reconstruction efficiency $\epsilon$ is defined as the fraction of MuCS-triggered cosmic-ray events that have a reconstructed track in the TPC:
\begin{equation}\label{eq:eff}
  \epsilon = \frac{\mathrm{reco.~MuCS~cosmic\myhyphen ray~events}}{\mathrm{MuCS~triggered~events}}=\frac{M_{e}}{T_{e}}.
\end{equation}
In data, the MuCS-tagged track is defined as the reconstructed track with the closest starting point $(x_{\mathrm{reco}},y_{\mathrm{reco}},z_{\mathrm{reco}})$ to the extrapolated MuCS starting point $(x_{\mathrm{MuCS}},y_{\mathrm{MuCS}},z_{\mathrm{MuCS}})$, as shown in figure~\ref{fig:evd}. 

In order to limit the accidental misassociation of MuCS-triggered cosmic-ray muons with other nearby reconstructed tracks in the TPC, a selection requirement is placed on the maximum distance $d_{\mathrm{max}}$ between the two points, $(x_{\mathrm{reco}},y_{\mathrm{reco}},z_{\mathrm{reco}})$ and $(x_{\mathrm{MuCS}},y_{\mathrm{MuCS}},z_{\mathrm{MuCS}})$.
To study the dependence of the number of MuCS-tagged tracks on $d_{\mathrm{max}}$, a dedicated Monte Carlo simulation of a MuCS run is performed, defined as \emph{MuCS Monte Carlo}, which is different from the Monte Carlo sample described in section \ref{sec:mcgen}. Each event of this simulation has one cosmic-ray muon passing through the MuCS boxes overlaid on a full simulation of cosmic rays in the TPC.

\begin{figure}[htbp]
  \begin{center}
    \includegraphics[width=0.7\linewidth]{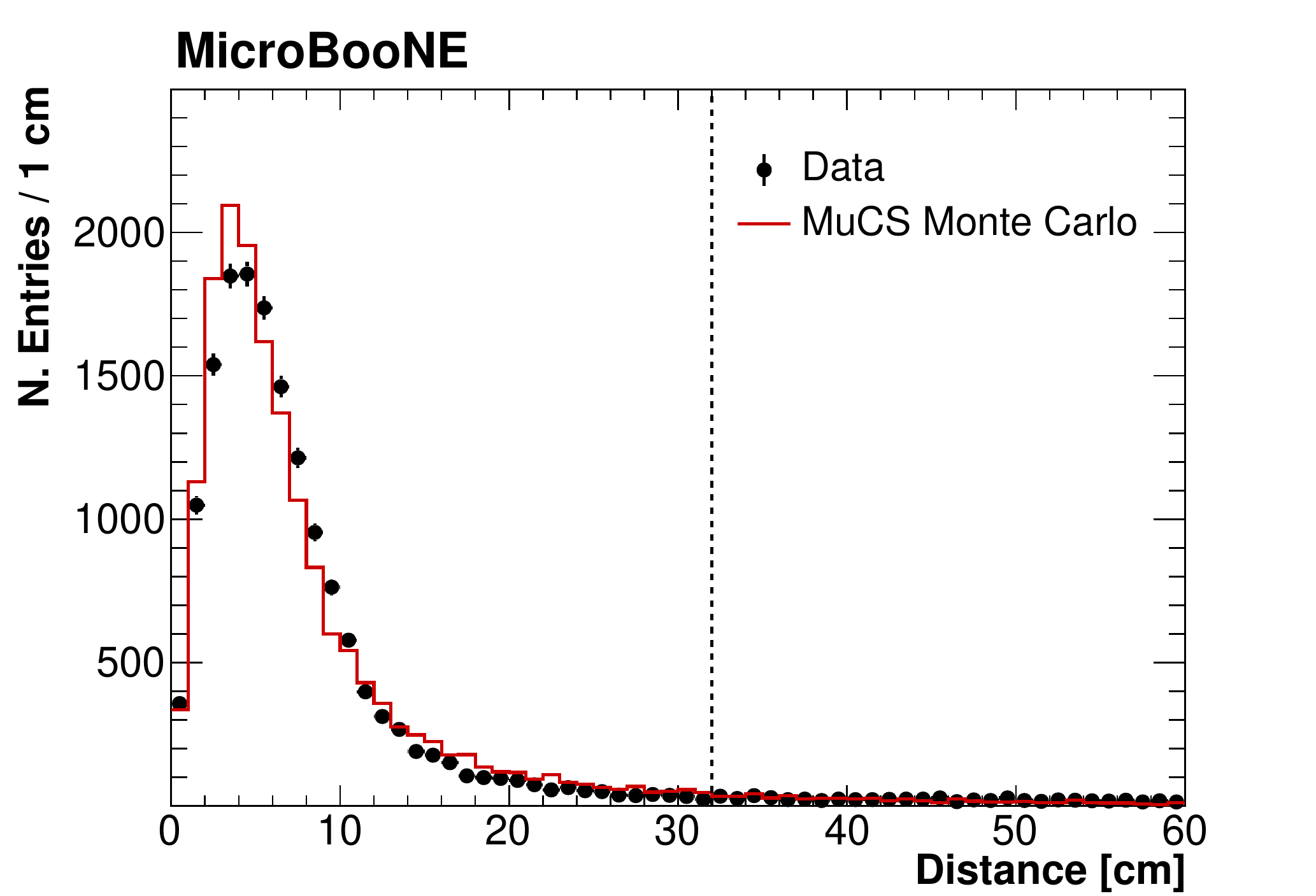}
    \caption{Data and MuCS Monte Carlo distributions of the distance $d$ between the extrapolated starting point and the closest reconstructed starting point for cosmic ray tracks passing through both the MuCS and the MicroBooNE TPC. The dashed line correspond to the $d_{\mathrm{max}}$ requirement (32 cm) chosen for this analysis.} \label{fig:dist}
  \end{center}
\end{figure}

We use the truth information in the MuCS Monte Carlo simulation to determine if the identified MuCS-tagged cosmic ray corresponds to the true track from the cosmic muon or if it is an incorrectly associated cosmic ray, to which the extrapolated starting point distance is closer than $d_{\mathrm{max}}$ because of multiple Coulomb scattering.
In the MuCS Monte Carlo the distance $d$ is defined as
\begin{equation}\label{eq:d_mc}
d = \sqrt{(x_{\mathrm{sim}}-x_{\mathrm{reco}})^2+(y_{\mathrm{sim}}-y_{\mathrm{reco}})^2+(z_{\mathrm{sim}}-z_{\mathrm{reco}})^2},
\end{equation}
where $(x_{\mathrm{sim}},y_{\mathrm{sim}},z_{\mathrm{sim}})$ and $(x_{\mathrm{reco}},y_{\mathrm{reco}},z_{\mathrm{reco}})$ are the coordinates of the intersection of the simulated cosmic-ray trajectory with the TPC and of the closest reconstructed track, respectively. Figure~\ref{fig:dist} shows the distribution of the distance $d$ between the extrapolated starting point and the closest reconstructed starting point, for both data and MuCS Monte Carlo.
In the reconstruction efficiency definition for the MuCS Monte Carlo sample, we replace the number of MuCS-triggered events in eq. \eqref{eq:eff} with the number of simulated MuCS events. 

The efficiency $\epsilon_{\mathrm{tag}}$ of the $d_{\mathrm{max}}$ requirement is defined as
\begin{equation}
  \epsilon_{\mathrm{tag}}=\frac{\mathrm{events~with~a~reco.~cosmic~ray~within~}d_{\mathrm{max}}}{\mathrm{MuCS~triggered~events}} = \frac{R_{e}(d_{\mathrm{max}})}{T_{e}}.
\end{equation}
It is calculated for the data sample ($\epsilon^{\mathrm{data}}_{\mathrm{tag}}$) and for the MuCS Monte Carlo sample ($\epsilon^{\mathrm{MuCS-MC}}_{\mathrm{tag}}$) by replacing the number of MuCS-triggered events with the number of simulated MuCS events.

The purity $P$ of the Monte Carlo MuCS sample, which represents the fraction of correctly tagged MuCS cosmic rays, is defined as the ratio between the number of events with a reconstructed MuCS cosmic ray correctly identified within $d_{\mathrm{max}}$ and the number of events with a reconstructed cosmic ray within $d_{\mathrm{max}}$ (MuCS-tagged cosmic rays):
\begin{equation}
  P=\frac{\mathrm{events~with~a~reco.~MuCS~cosmic~ray~within~}d_{\mathrm{max}}}{\mathrm{events~with~a~reco.~cosmic~ray~within~}d_{\mathrm{max}}} = \frac{M_{e}(d_{\mathrm{max}})}{R_{e}(d_{\mathrm{max}})}.
\end{equation}
The acceptance $A$ of the $d_{\mathrm{max}}$ requirement, which represents the portion of MuCS cosmic rays within $d_{\mathrm{max}}$, is defined as the ratio between the number of events with a reconstructed MuCS cosmic ray within $d_{\mathrm{max}}$ range and the total number of events with a reconstructed MuCS cosmic ray:
\begin{equation}
  A=\frac{\mathrm{events~with~a~reco.~MuCS~cosmic~ray~within~}d_{\mathrm{max}}}{\mathrm{events~with~a~reco.~MuCS~cosmic~ray}} = \frac{M_{e}(d_{\mathrm{max}})}{M_{e}}.
\end{equation}
The acceptance of the $d_{\mathrm{max}}$ requirement is mainly affected by the multiple Coulomb scattering in the material between the MuCS and the TPC.

The reconstruction efficiency, as defined in eq. \eqref{eq:eff}, is obtained, both for data ($\epsilon_{\mathrm{data}}$) and for MuCS Monte Carlo ($\epsilon_{\mathrm{MuCS\myhyphen MC}}$), by
\begin{equation}\label{eq:mceff}
  \epsilon = \frac{M_{e}}{T_{e}} = \frac{R_{e}(d_{\mathrm{max}})}{T_{e}} \times \frac{M_{e}(d_{\mathrm{max}})}{R_{e}(d_{\mathrm{max}})} \times \frac{M_{e}}{M_{e}(d_{\mathrm{max}})} = \epsilon_{\mathrm{tag}} \times \frac{P}{A},
\end{equation}
where the $P/A$ ratio is taken only from the MuCS Monte Carlo simulation, while $\epsilon_{\mathrm{tag}}$ is measured with the data, $\epsilon^{\mathrm{data}}_{\mathrm{tag}}$, or with the MuCS Monte Carlo simulation, $\epsilon^{\mathrm{MuCS\myhyphen MC}}_{\mathrm{tag}}$.

\begin{figure}[htbp]
  \begin{center}
    \includegraphics[width=0.7\linewidth]{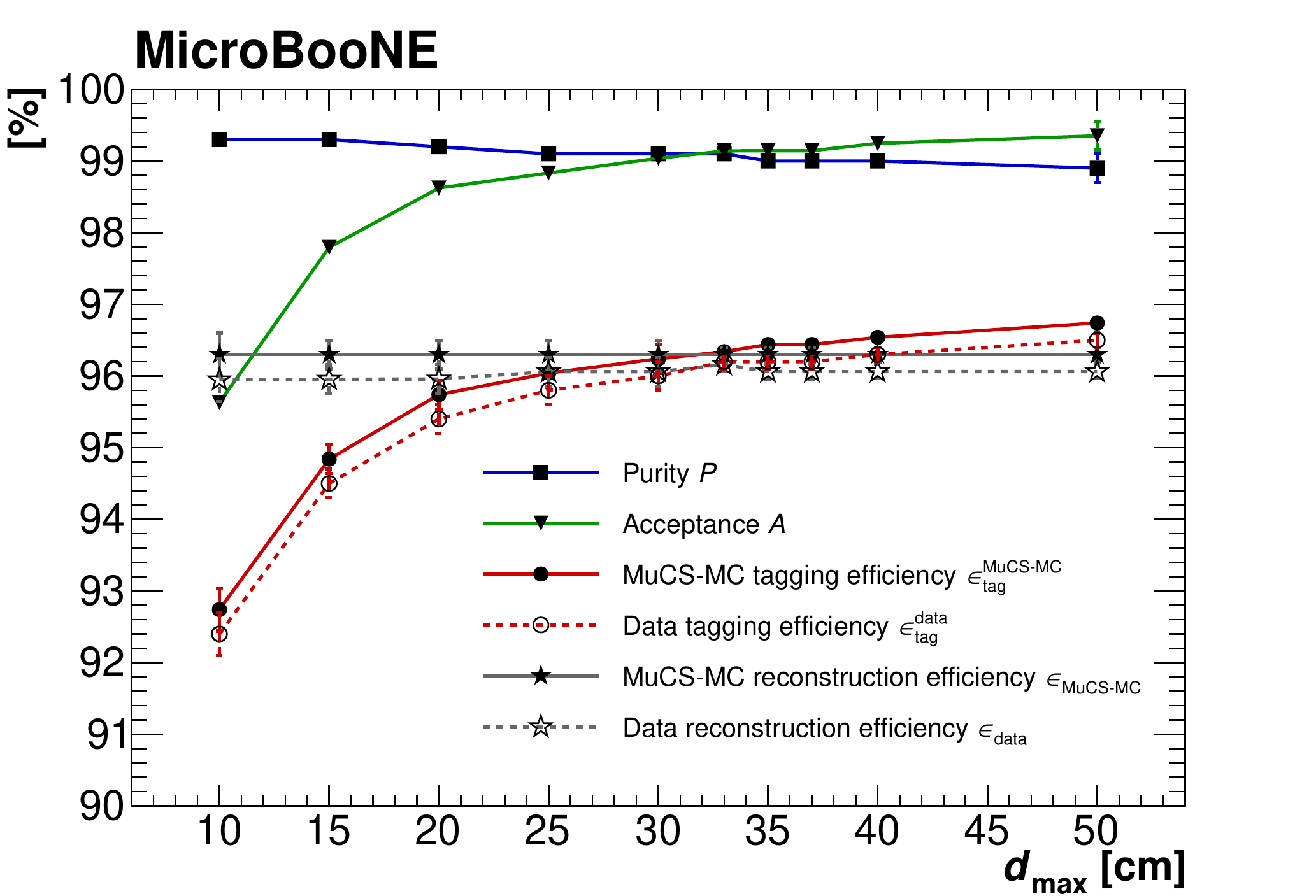}
    \caption{Data ($\epsilon^{\mathrm{data}}_{\mathrm{tag}}$) and Monte Carlo ($\epsilon^{\mathrm{MC}}_{\mathrm{tag}}$) tagging efficiency (red), purity $P$ (blue) and acceptance $A$ (green), as a function of $d_{\mathrm{max}}$. The MuCS reconstruction efficiencies for data ($\epsilon_{\mathrm{data}}$) and MuCS Monte Carlo ($\epsilon_{\mathrm{MuCS\myhyphen MC}}$) are also shown as a reference (grey).} \label{fig:purity}
  \end{center}
\end{figure}

Figure~\ref{fig:purity} shows the tagging efficiency both for data ($\epsilon^{\mathrm{data}}_{\mathrm{tag}}$) and MuCS Monte Carlo  ($\epsilon^{\mathrm{MuCS\myhyphen MC}}_{\mathrm{tag}}$), the purity $P$ and the acceptance $A$ as a function of $d_{\mathrm{max}}$. The reconstruction efficiencies for data ($\epsilon_{\mathrm{data}}$) and MuCS Monte Carlo ($\epsilon_{\mathrm{MuCS\myhyphen MC}}$) are also shown.

Using eq. \eqref{eq:eff}, the MuCS Monte Carlo reconstruction efficiency $\epsilon_{\mathrm{MuCS\myhyphen MC}}$ will not depend, by construction, on the chosen value of $d_{\mathrm{max}}$. Since the $P/A$ correction factor is determined by a Monte Carlo simulation, the data reconstruction efficiency $\epsilon_{\mathrm{data}}$ has a small dependence on $d_{\mathrm{max}}$ (see figure~\ref{fig:purity}), because of the small difference between $\epsilon^{\mathrm{data}}_{\mathrm{tag}}$ and $\epsilon^{\mathrm{MuCS\myhyphen MC}}_{\mathrm{tag}}$.
The difference between the lowest and the highest value of $\epsilon_{\mathrm{data}}$ is $0.2\%$. This value is used to estimate the systematic uncertainty related to the $P/A$ correction factor, as further discussed in section \ref{sec:sys}.
The chosen value of $d_{\mathrm{max}}$ is $d_{\mathrm{max}}=32~\mathrm{cm}$. Figure~\ref{fig:dist} and figure~\ref{fig:purity} show that the $d$, $P$ and $A$ distributions are constant around this value, and the ratio $P/A$ is $\approx$1.



To verify if the data reconstruction efficiency, measured in specific locations of the detector, is valid throughout the detector, we perform a direct comparison of the MuCS data with the Monte Carlo sample, using the Monte Carlo distribution generated as described in section \ref{sec:mcgen}.
The Monte Carlo cosmic-ray reconstruction efficiency is defined as
\begin{equation}
  \epsilon_{\mathrm{MC}} = \frac{\mathrm{reco.~cosmic\myhyphen ray~tracks}}{\mathrm{generated~cosmic~rays}}.
\end{equation}
This Monte Carlo sample contains cosmic rays generated over the entire TPC volume and therefore averages over any dependence of $\epsilon_{\mathrm{MC}}$ on the position of the cosmic ray in the TPC. The MuCS dataset, however, covers only three regions of the detector shown in figure~\ref{fig:mucs}. 

A cosmic ray with a longer path in the TPC will correspond in general to a larger number of hit wires and thus to a higher reconstruction efficiency \cite{pandora2}. The reconstruction efficiency depends also on the direction of the cosmic ray, since cosmic rays parallel to the wires of one plane ($0^{\circ}$, $\pm60^{\circ}$ with respect to the $y$ axis) will generate fewer hits in that particular plane, making the track reconstruction more difficult.
We therefore express the data and the Monte Carlo reconstruction efficiencies, $\epsilon_{\mathrm{data}}$ and $\epsilon_{\mathrm{MC}}$, as a function of the starting spherical angles $\theta$, $\phi$ and of the expected track length $L$ in the TPC, as described in section~\ref{sec:merging}.

The efficiency is plotted as a three-dimensional histogram, where each bin corresponds to a particular combination of the $\theta$, $\phi$, and $L$ variables. Bin width is chosen large enough to have a statistical uncertainty of less than $10\%$ for every ($\theta$, $\phi$, $L$) bin. The same $P/A$ correction factor is applied to every bin.

The data reconstruction efficiency $\epsilon_{\mathrm{data}}$ does not take into account muons triggering the MuCS that decay or are captured before reaching the TPC. These events are counted in the denominator of eq. \eqref{eq:eff} and therefore lower the reconstruction efficiency since they have not reached the TPC and cannot be reconstructed. The fraction $D$ of cosmic rays that traverse the MuCS but decay or are captured before reaching the TPC is measured from the MuCS Monte Carlo simulation as

\begin{equation}
D = \frac{\mathrm{decayed/captured~muons}}{\mathrm{MuCS~triggered~events}} = (1.0 \pm 0.1)\%.
\end{equation}
This correction factor does not show a dependence of $\theta$, $\phi$, and $L$ with the present level of statistics. The corrected data reconstruction efficiency is given by
\begin{equation}\label{eq:dcorr}
\epsilon_{\mathrm{data}}^{\mathrm{corr}} =  \frac{\epsilon_{\mathrm{data}}}{1-D}.
\end{equation}

\begin{figure}[htbp]
  \begin{subfigure}{0.495\textwidth}
    \includegraphics[width=\linewidth]{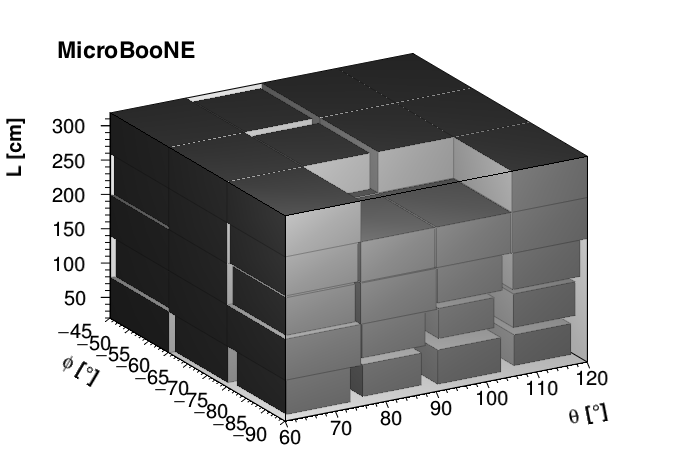}
    \caption{Monte Carlo} \label{fig:3d_mc}
  \end{subfigure}
  \begin{subfigure}{0.495\textwidth}
    \includegraphics[width=\linewidth]{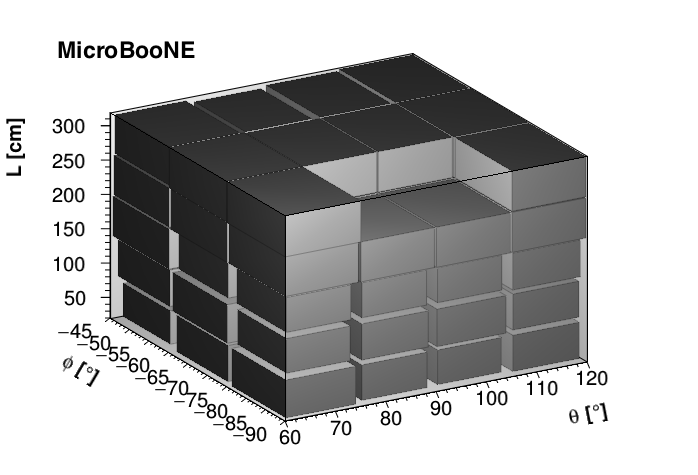}
    \caption{Data} \label{fig:3d_data}
  \end{subfigure}
  \caption{Three-dimensional representation of reconstruction efficiency as a function of the starting angles $\theta$, $\phi$ and the extrapolated track length $L$ for (a) Monte Carlo and (b) data. The size of the box represents the efficiency. The empty region in the upper part of the plot corresponds to a region of the parameter space not covered by the data sample.}\label{fig:3d}
\end{figure}

Figure~\ref{fig:3d} shows both the corrected data and the Monte Carlo reconstruction efficiency in the 3D phase space $\theta, \phi, L$, calculated as described in section \ref{sec:merging}. The average reconstruction efficiencies for the data and Monte Carlo samples considering only statistical uncertainties, are
\begin{align}
\epsilon_{\mathrm{data}}^{\mathrm{corr}} &= (97.1 \pm 0.1)\%\\
\epsilon_{\mathrm{MC}} &= (97.4 \pm 0.1)\% \nonumber
\end{align}
for data and Monte Carlo, respectively.

\subsection{Systematic uncertainties}\label{sec:sys}
The measurement of the reconstruction efficiency can be affected by several systematic effects. In particular, the datasets are taken with the MuCS placed in three different geometrical configurations and the MuCS-triggered cosmic muons undergo multiple Coulomb scattering. In this section, the details of all the systematic effects studied are given.

\subsubsection{Effect of the space-charge effect correction}
The displacement of the ionization tracks caused by the space-charge effect has been corrected with the procedure described in section \ref{sec:merging}. The error related to this correction is verified to have a negligible effect ($<0.1\%$) on the data reconstruction efficiency measurement.

\subsubsection{Effect of the $d_{\mathrm{max}}$ requirement}

As shown in figure~\ref{fig:purity}, the value of $\epsilon_{\mathrm{data}}$ has a small dependence on $d_{\mathrm{max}}$. The difference $0.2\%$ between the highest and lowest value of $\epsilon_{\mathrm{data}}$ is taken as the systematic uncertainty due to the $d_{\mathrm{max}}$ requirement. This difference could be caused by the multiple Coulomb scattering of the cosmic rays in the box material, which is not included in the MuCS simulation.

\subsubsection{Decay-in-flight or captured muons}\label{sec:dif}
The Monte Carlo statistical uncertainty of the correction factor $D$, as defined in eq. \eqref{eq:dcorr}, is 0.1\% and is taken as the systematic uncertainty related to this correction.

\subsubsection{Detector non-uniformities}\label{sec:wires}
The presence of potential detector non-uniformities can introduce a systematic uncertainty in the measurement of the reconstruction efficiency. In particular, the presence of noisy or unresponsive wires in specific regions of the detector can lower the reconstruction efficiency. The three different MuCS configurations (shown in figure~\ref{fig:mucs}) cover different regions of the TPC, providing information on potential non-uniformities.

To check if these non-uniformities introduce a systematic effect, the significance $\sigma$ of the differences between the data reconstruction efficiencies measured for two different configurations with the following definition is calculated as
\begin{equation}
\sigma = \frac{\epsilon_a-\epsilon_b}{\sqrt{(\Delta \epsilon_{a})^2 + (\Delta \epsilon_b)^2}},
\end{equation}
where $\epsilon_{a}$ ($\epsilon_{b}$) is the reconstruction efficiency in the arbitrary $a$ ($b$) configuration studied and $\Delta \epsilon_{a}$ ($\Delta \epsilon_{b}$) is the corresponding statistical uncertainty. This significance is measured for each corresponding ($\theta,\phi,L$) bin and for each possible combination of central, downstream, and upstream configurations. In the presence of a systematic effect, the standard deviation of the significances distribution would be larger than unity. A Gaussian fit of the distribution gives a standard deviation of $1.54\pm0.12$, suggesting that detector non-uniformities are indeed present.

\begin{figure}[htbp]
  \begin{center}
    \begin{subfigure}{0.48\textwidth}
      \includegraphics[width=\linewidth]{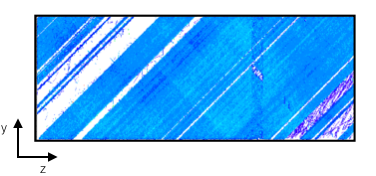}
      \caption{Collected charge}
    \end{subfigure}
    \begin{subfigure}{0.48\textwidth}
      \includegraphics[width=\linewidth]{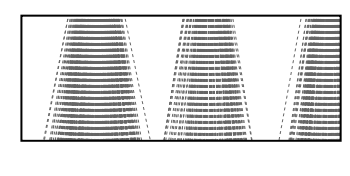}
      \caption{Extrapolated tracks}
    \end{subfigure}

    \caption{(a): two-dimensional display of the collected charge in the MicroBooNE detector, showing the regions with missing charges due to noisy or unresponsive wires on one of the induction planes. (b): extrapolated tracks corresponding to bins with $\sigma>3$, as described in the text. As shown in (b), the tracks in the upstream part of the detector (low $z$) traverse a region with several noisy or unresponsive wires.} \label{fig:wires}
  \end{center}
\end{figure}

\begin{figure}[htbp]
  \begin{center}
    \begin{subfigure}{0.5\textwidth}
      \includegraphics[width=\linewidth]{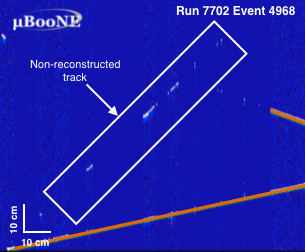}
      \caption{Induction ($+60^{\circ}$) plane} \label{fig:u}
    \end{subfigure}
    \begin{subfigure}{0.5\textwidth}
      \includegraphics[width=\linewidth]{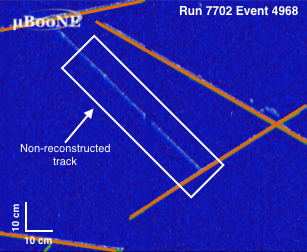}
      \caption{Induction ($-60^{\circ}$) plane} \label{fig:v}
    \end{subfigure}
    \begin{subfigure}{0.5\textwidth}
      \includegraphics[width=\linewidth]{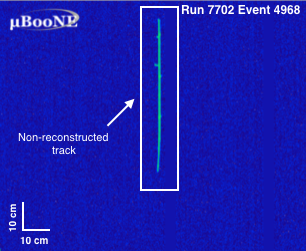}
      \caption{Collection ($0^{\circ}$) plane} \label{fig:y}
    \end{subfigure}    \caption{Event display of a non-reconstructed MuCS cosmic-ray track in the three wire planes, indicated in the white boxes. Orange lines correspond to other TPC reconstructed tracks. The collection plane shows no reconstructed tracks.} \label{fig:example}
  \end{center}
\end{figure}

We study cosmic rays corresponding to bins with larger $\sigma$ in more detail and two contributing factors that lead to non-uniformities: regions of the detector with unresponsive wires and highly inclined tracks. The MuCS-extrapolated tracks from the bins with a $\sigma>3$, which drive the broadening of the distribution, show that for the upstream configuration the cosmic rays go through regions with noisy or unresponsive wires in one of the induction planes (figure~\ref{fig:wires}), which are the source of the detector non-uniformities.
In addition, the MuCS-extrapolated tracks in these bins have an orientation as shown in figure~\ref{fig:wires}, implying that the cosmic rays are aligned with the wires of the collection plane, parallel to the $y$ axis.  These tracks, therefore, have few hits in two of the three planes, affecting the reconstruction efficiency. Figure~\ref{fig:example} shows an event display of a non-reconstructed MuCS cosmic ray going through the region with noisy or unresponsive wires and parallel to the collection plane wires.

The systematic uncertainty related to the detector non-uniformities for each $\theta,\phi,L$ bin is calculated as the difference between the best reconstruction efficiency of the three configurations and the averaged reconstruction efficiency obtained by merging the three datasets.
The systematic uncertainty for the integrated 3D efficiency is $1.1\%$.

\subsubsection{Energy sampling}
The multiple Coulomb scattering of cosmic muons depends on the energy of the cosmic ray \cite{pdg}. Cosmic rays that scatter more have a higher probability to be outside the $d_{\mathrm{max}}$ region and they are also more difficult to reconstruct, since their path in the TPC is not a straight line. In the ($\theta$, $\phi$, $L$) bins where the data reconstruction efficiency is measured with low statistics, the MuCS cosmic rays can be distributed in a small region of the energy spectrum, which is steeply falling~\cite{corsika}. This systematic bias of the reconstruction efficiency, estimated with a dedicated Monte Carlo simulation, is found to be negligible with the present level of statistics ($<0.1\%$).

\subsection{Data/Monte Carlo comparison}\label{sec:datamc}
The reconstruction efficiencies for the Monte Carlo and data samples are calculated as described in section~\ref{sec:reco}. 

\begin{figure}[htbp]
  \begin{subfigure}{0.32\textwidth}
    \includegraphics[width=\linewidth]{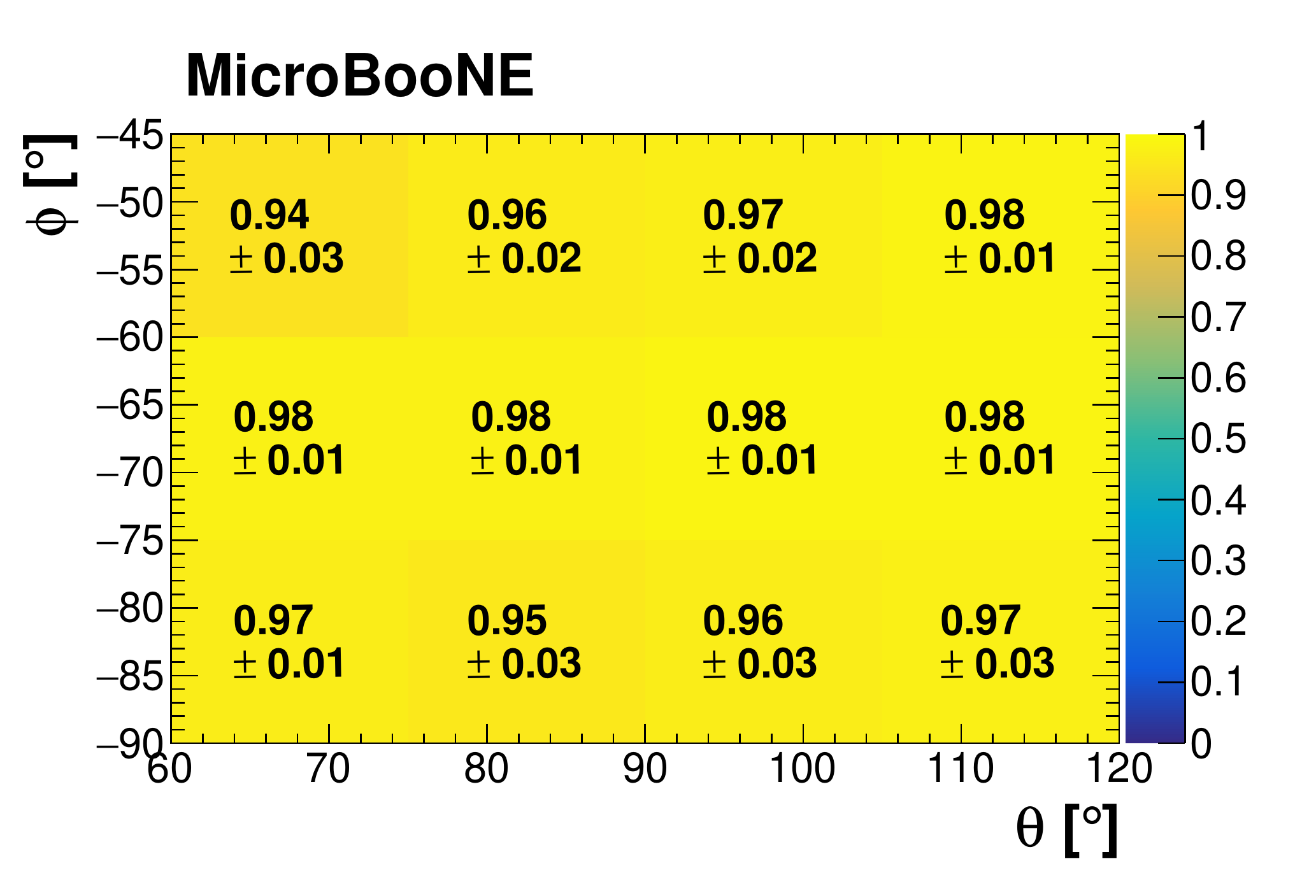}
    \caption{$(\theta,\phi)$ - data}
  \end{subfigure}\begin{subfigure}{0.32\textwidth}
  \includegraphics[width=\linewidth]{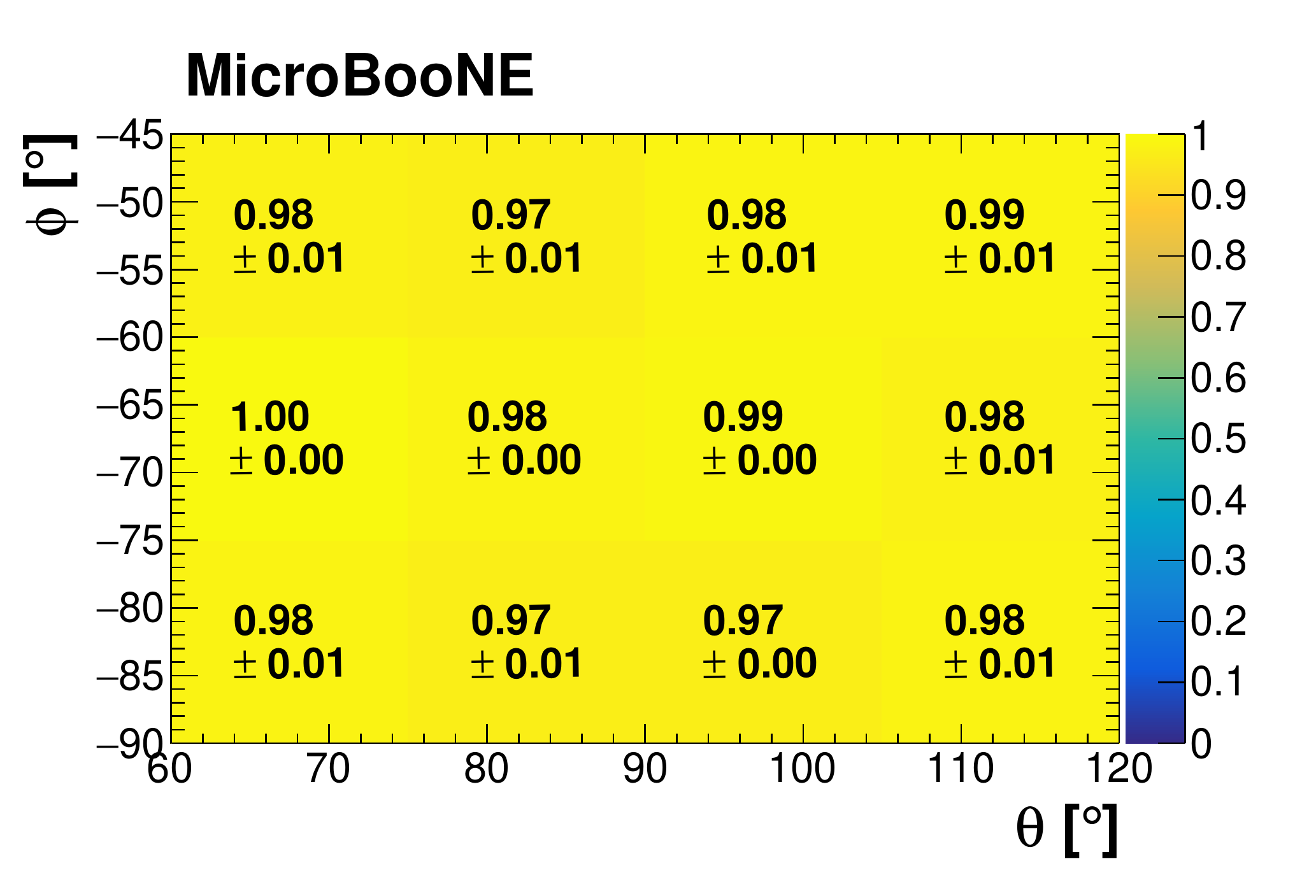}
  \caption{$(\theta,\phi)$ - Monte Carlo}
\end{subfigure}\begin{subfigure}{0.32\textwidth}
  \includegraphics[width=\linewidth]{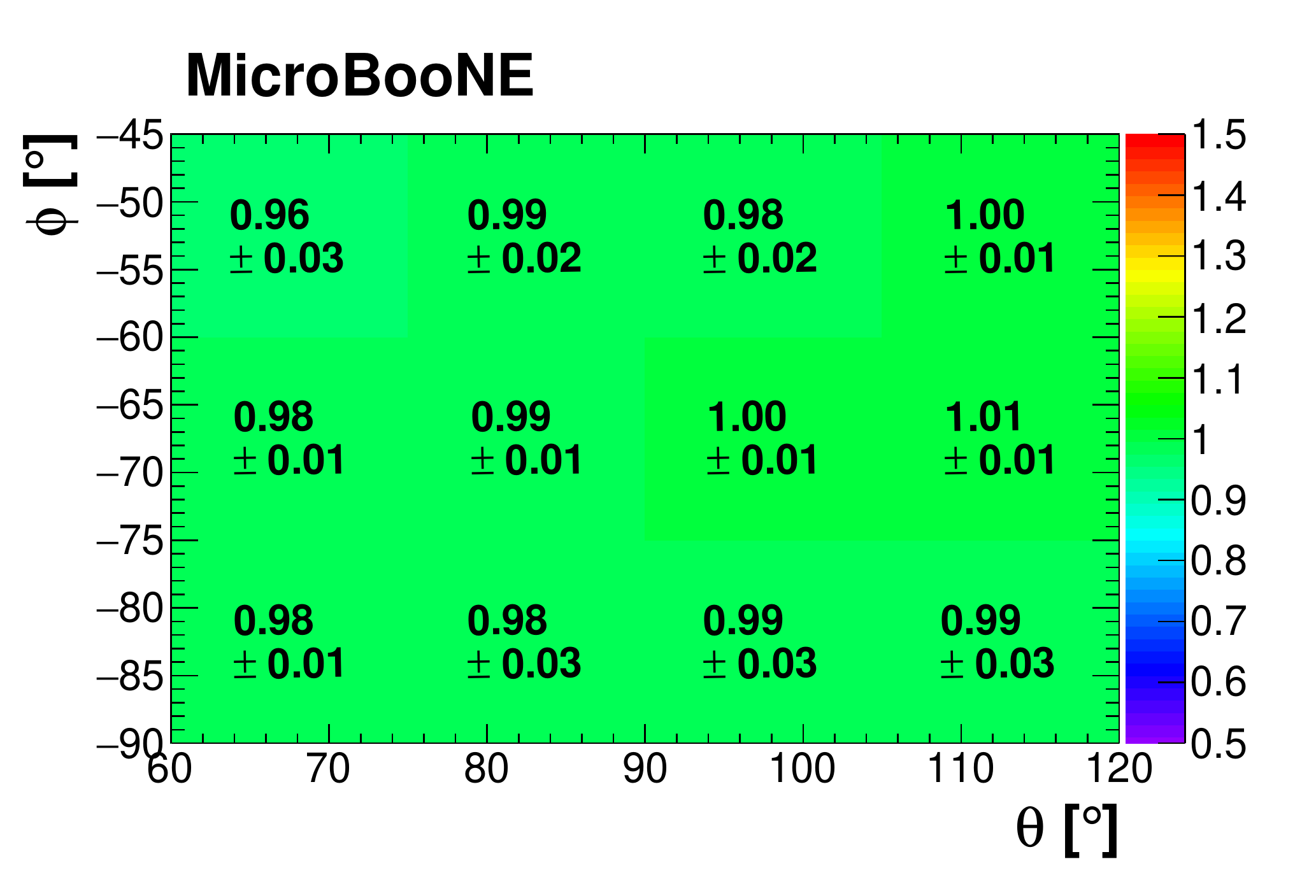}
  \caption{$(\theta,\phi)$ - data/Monte Carlo}
\end{subfigure}
\begin{subfigure}{0.32\textwidth}
  \includegraphics[width=\linewidth]{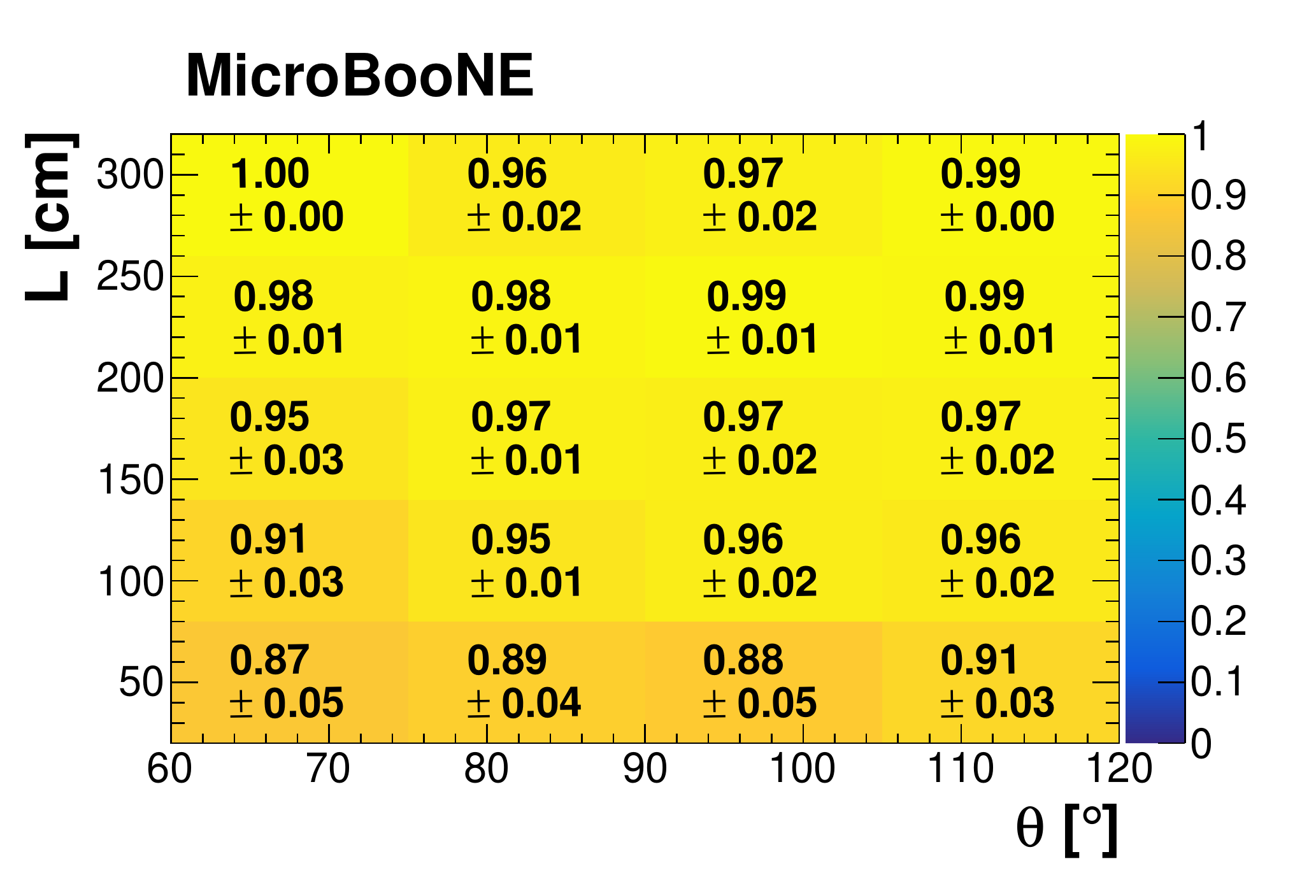}
  \caption{$(\theta,L)$ - data}
\end{subfigure}\begin{subfigure}{0.32\textwidth}
\includegraphics[width=\linewidth]{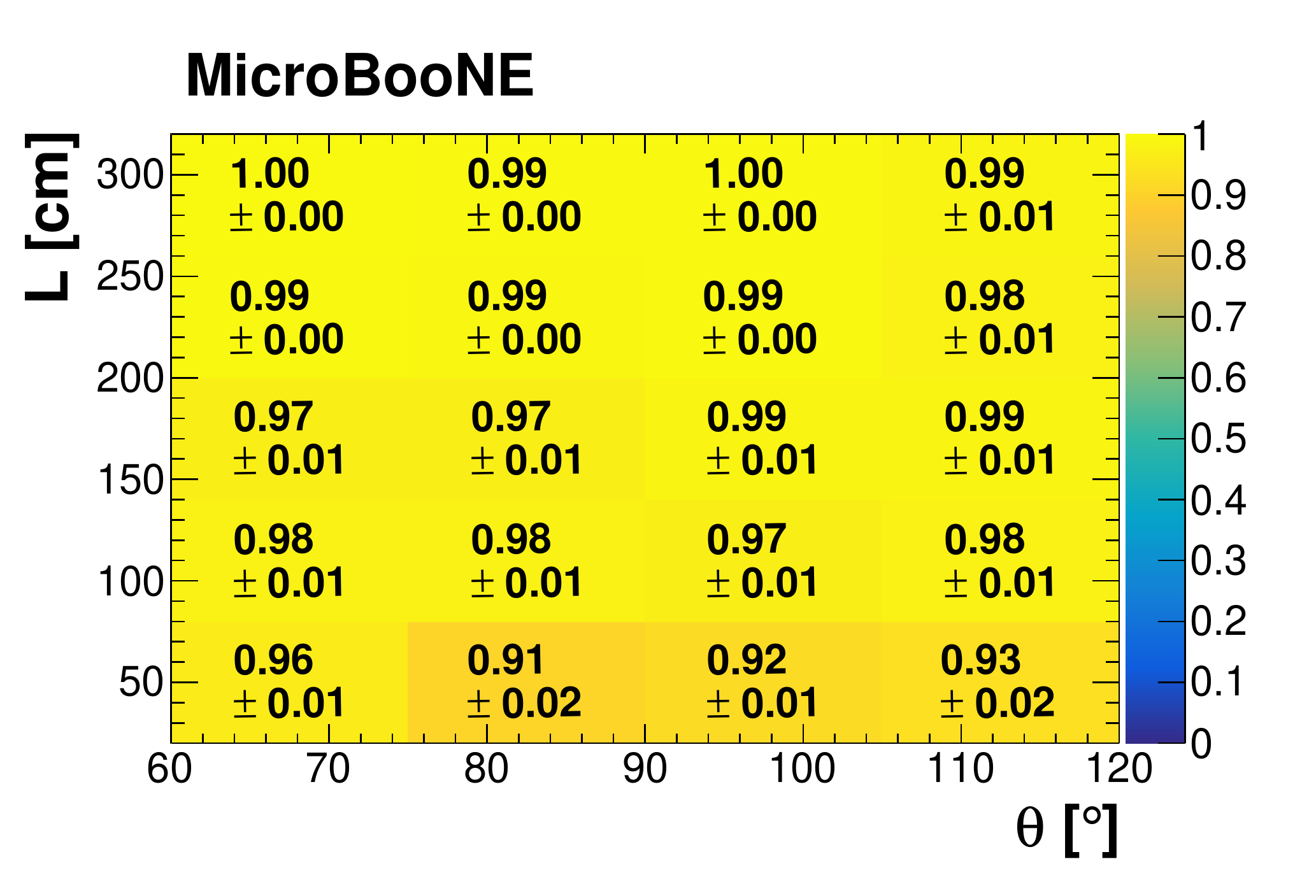}
\caption{$(\theta,L)$ - Monte Carlo}
\end{subfigure}\begin{subfigure}{0.32\textwidth}
\includegraphics[width=\linewidth]{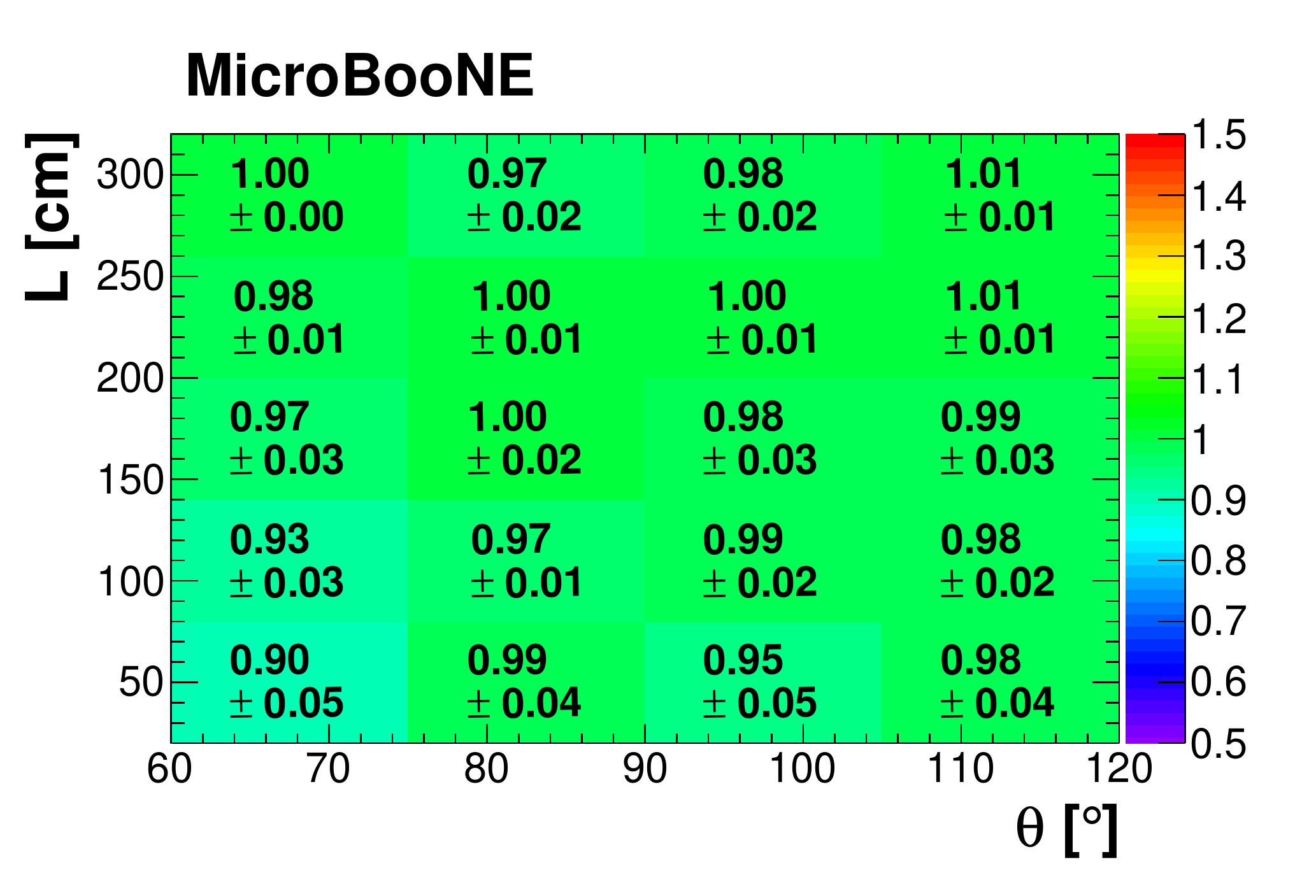}
\caption{$(\theta,L)$ - data/Monte Carlo}
\end{subfigure}
\begin{subfigure}{0.32\textwidth}
  \includegraphics[width=\linewidth]{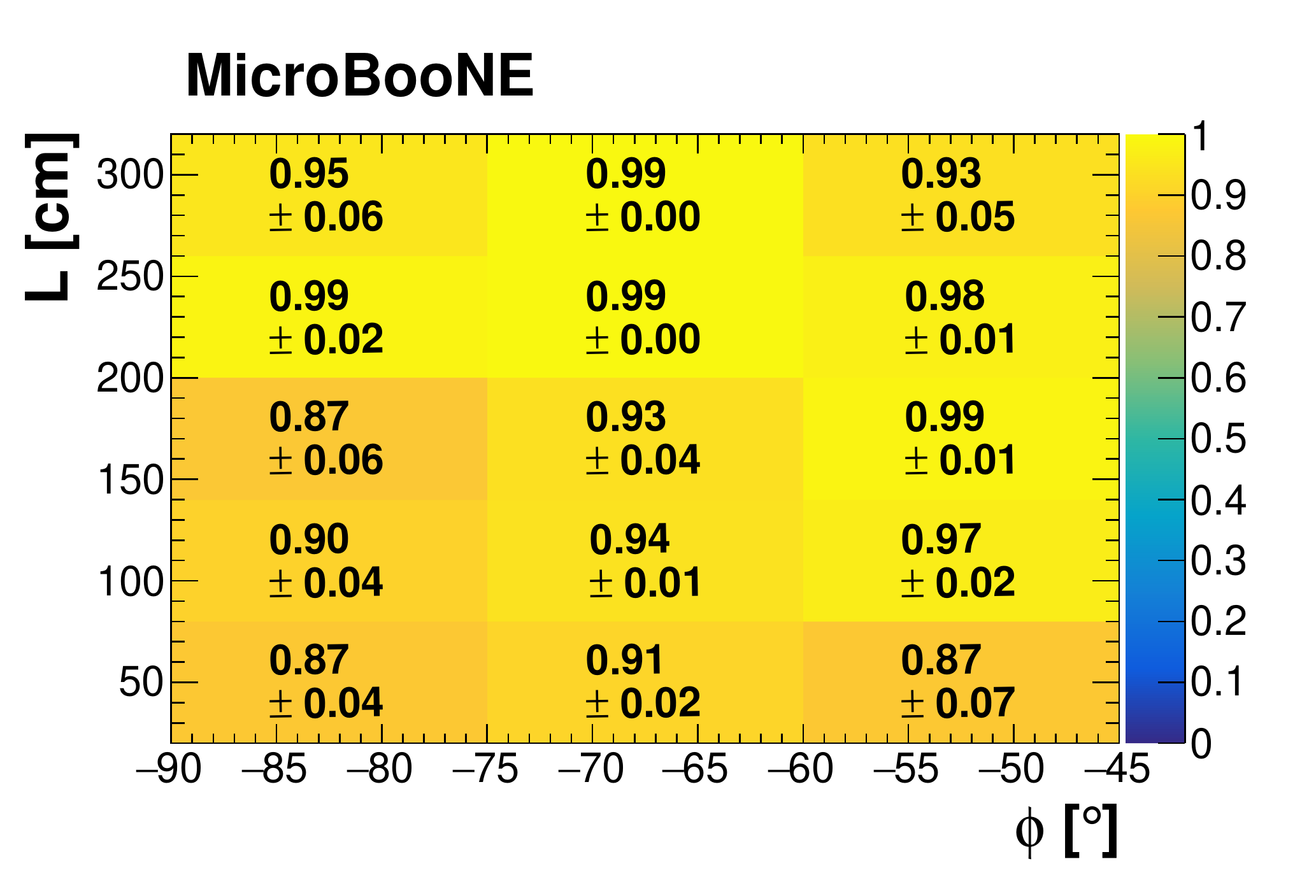}
  \caption{$(\phi,L)$ - data}
\end{subfigure}\begin{subfigure}{0.32\textwidth}
\includegraphics[width=\linewidth]{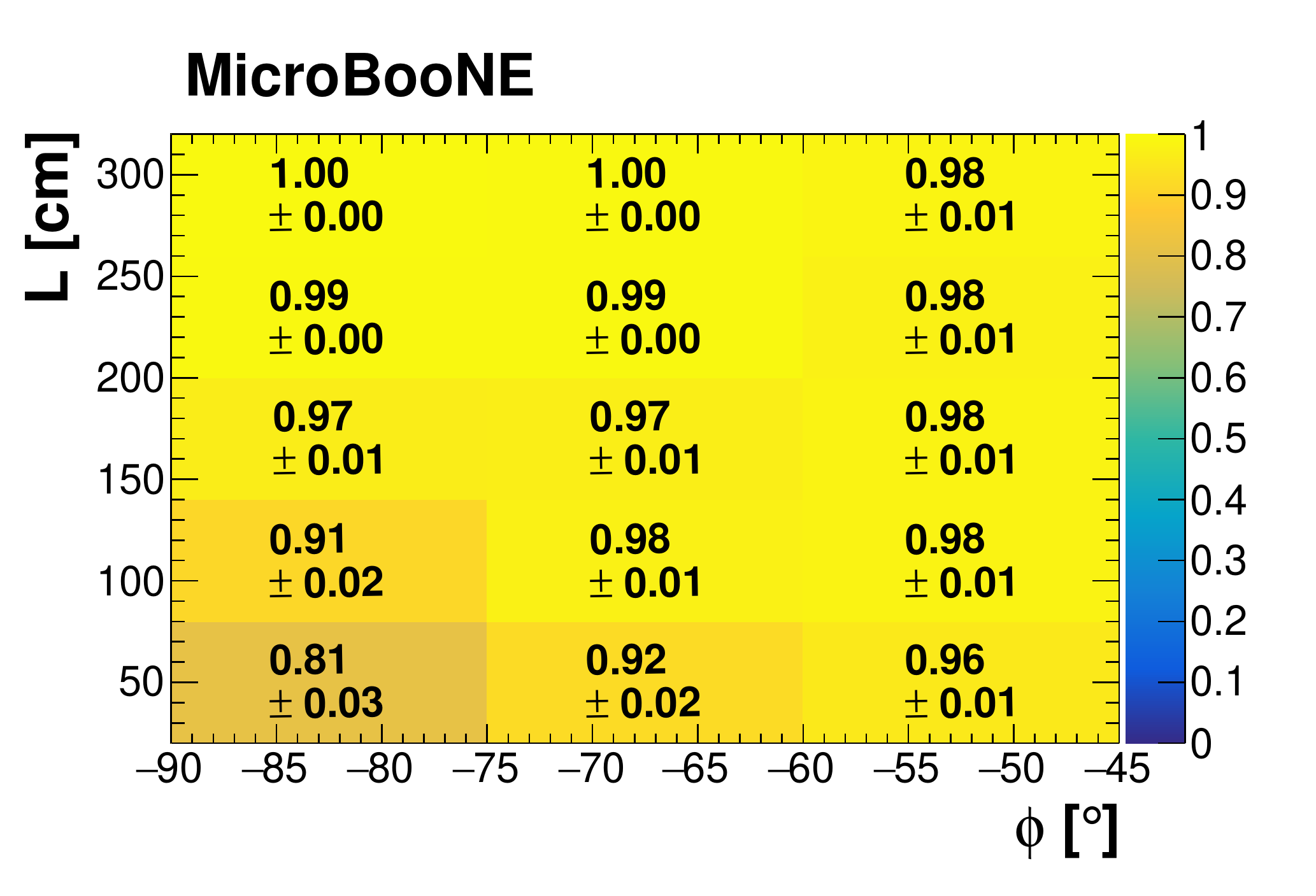}
\caption{$(\phi,L)$ - Monte Carlo}
\end{subfigure}\begin{subfigure}{0.32\textwidth}
\includegraphics[width=\linewidth]{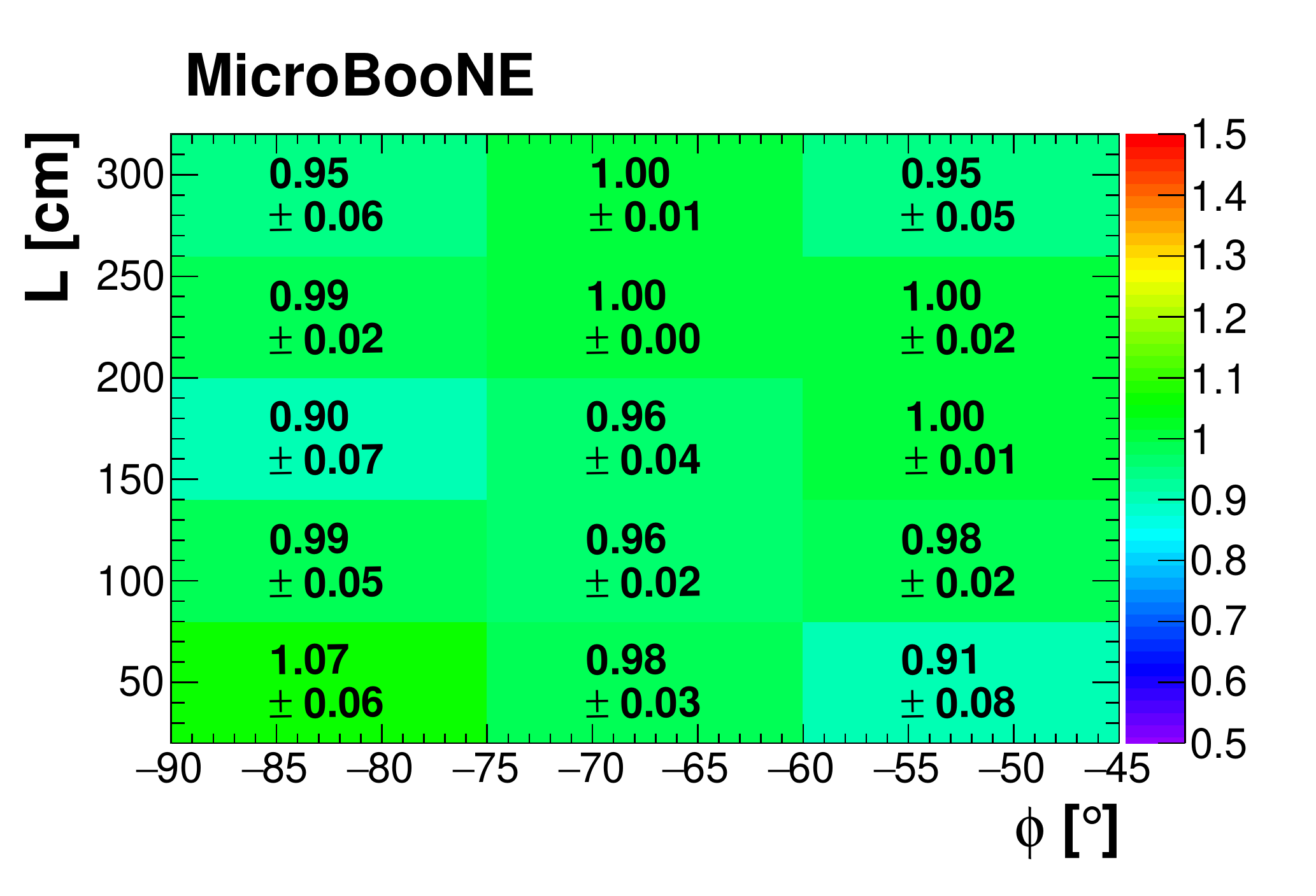}
\caption{$(\phi,L)$ - data/Monte Carlo}
\end{subfigure}
\caption{Two-dimensional representation of reconstruction efficiencies for data (left), Monte Carlo (center) and their ratio (right). Data uncertainties include systematic effects, while Monte Carlo uncertainties are statistical-only.}\label{fig:2d}
\end{figure}

\begin{figure}[htbp]
  \begin{center}
    \begin{subfigure}{0.61\textwidth}
      \includegraphics[width=\linewidth]{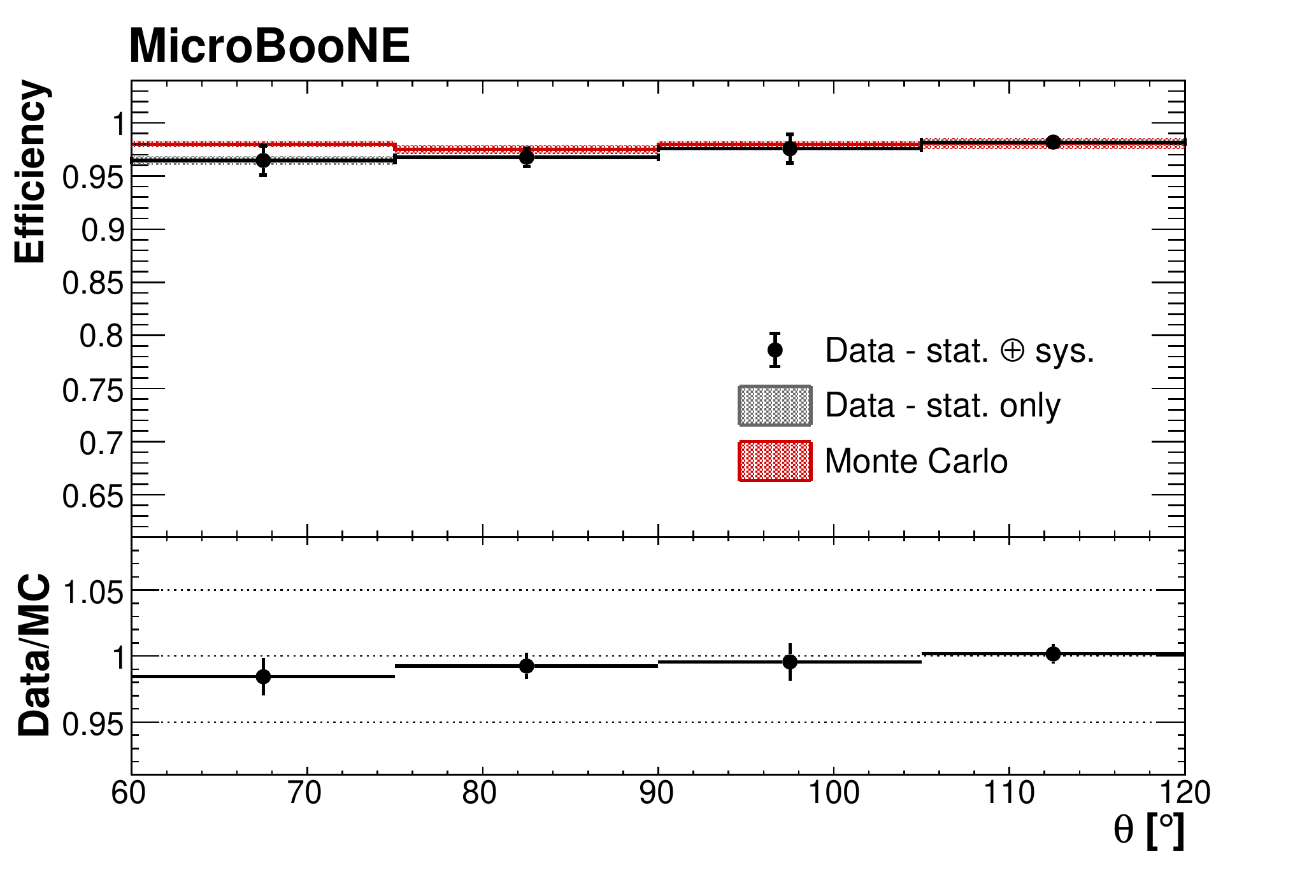}
      \caption{$\theta$} \label{fig:theta}
    \end{subfigure}
    \begin{subfigure}{0.61\textwidth}
      \includegraphics[width=\linewidth]{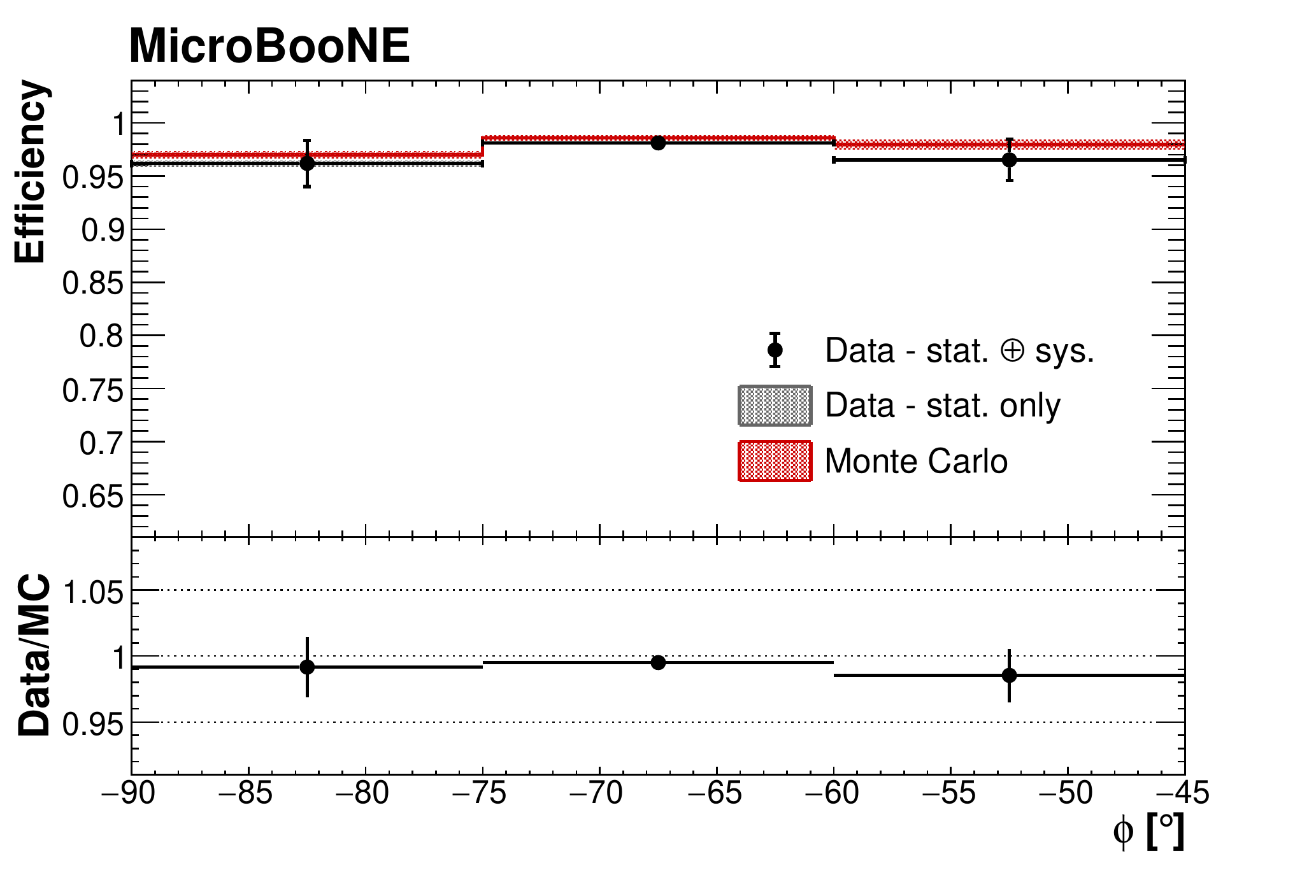}
      \caption{$\phi$} \label{fig:phi}
    \end{subfigure}
    \begin{subfigure}{0.61\textwidth}
      \includegraphics[width=\linewidth]{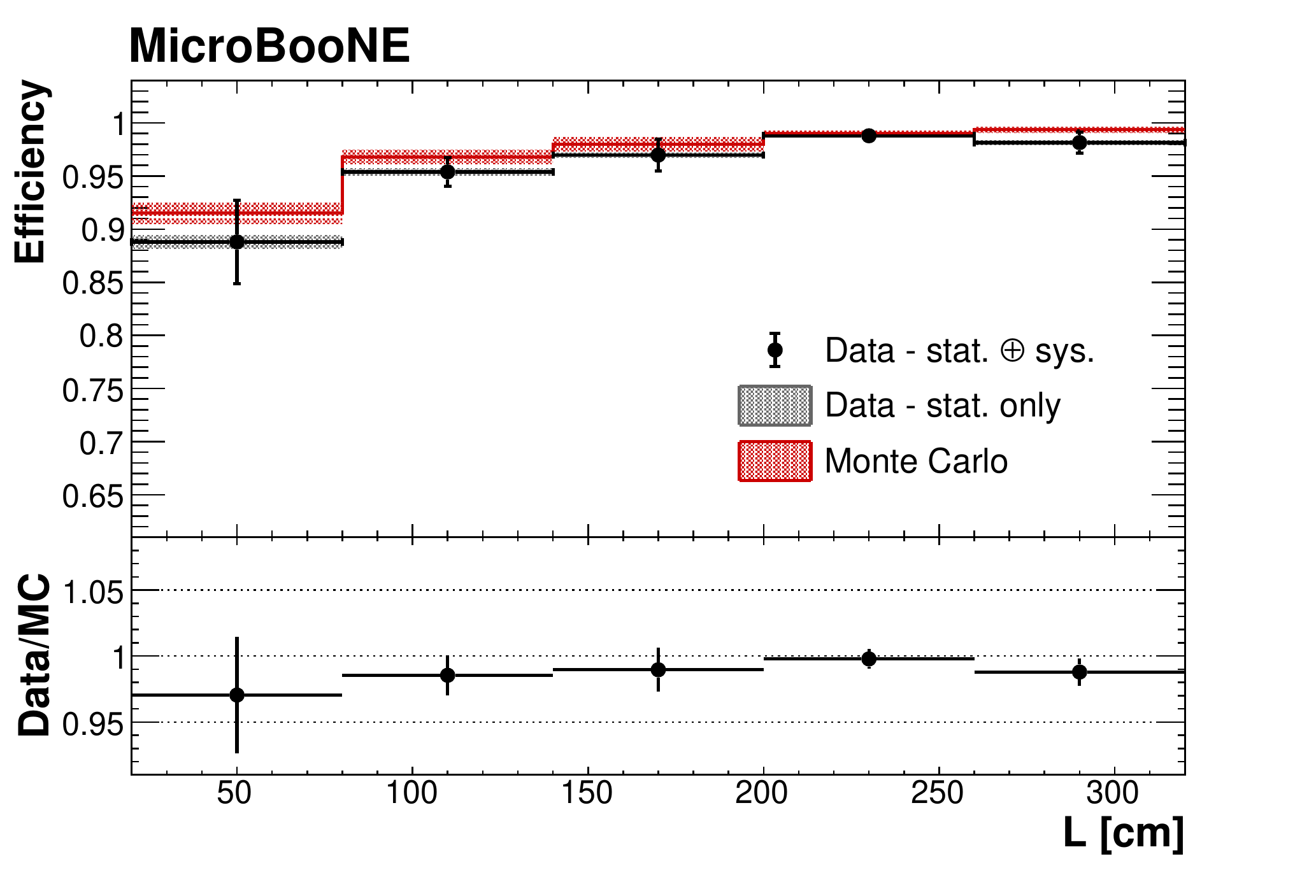}
      \caption{$L$} \label{fig:l}
    \end{subfigure}
    \caption{Monte Carlo (red line) and data (black points) reconstruction efficiency as a function of the starting angles $\theta$, $\phi$ and the extrapolated track length $L$. Data uncertainty bars include statistical uncertainties and systematic effects, while Monte Carlo uncertainties are statistical-only.}\label{fig:1d}
  \end{center}
\end{figure}

Figure~\ref{fig:2d} shows the efficiencies computed for the two-dimensional planes $(\theta,\phi)$, $(\theta,L)$, and $(\phi,L)$, and figure~\ref{fig:1d} shows the efficiencies computed as a function of $\theta$, $\phi$, and $L$ individually.

The reconstruction efficiency increases with the expected track length $L$ in the TPC, since longer tracks correspond, in general, to a larger number of hit wires that are easier to reconstruct.
The average reconstruction efficiencies for the data and Monte Carlo samples considering the systematic uncertainties, added in quadrature, in the analysis of the data are
\begin{align}
\epsilon_{\mathrm{data}}^{\mathrm{corr}} &= (97.1 \pm 0.1~\mathrm{(stat)} \pm 1.4~\mathrm{(sys)})\%,\\
\epsilon_{\mathrm{MC}} &= (97.4 \pm 0.1~\mathrm{(stat)})\%,\nonumber
\end{align} for data and Monte Carlo, showing good agreement within uncertainties.

\section{Conclusions}
Cosmic muons traversing a LArTPC detector located on the surface can produce a source of backgrounds to the analyses of neutrino interactions. Measuring the reconstruction efficiency of such cosmic rays in the detector is of fundamental importance for the assessment of the detector performance and the suppression of cosmic-ray background.

We present results using data from a small muon counter (the MuCS), placed above the MicroBooNE TPC, to measure the data reconstruction efficiency and compare it with the Monte Carlo reconstruction efficiency.
A method to evaluate the number of reconstructed MuCS cosmic rays is studied using a dedicated Monte Carlo simulation.

The data reconstruction efficiency, calculated by comparing the number of MuCS-triggered events with the number of events with a reconstructed MuCS cosmic ray, is measured as a function of the cosmic-ray starting angles $\theta$, $\phi$ and the expected length in the TPC, $L$. The overall reconstruction efficiency obtained is $\epsilon_{\mathrm{data}}=(97.1 \pm 0.1~\mathrm{(stat)} \pm 1.4~\mathrm{(sys)})\%$ and $\epsilon_{\mathrm{MC}} = (97.4 \pm 0.1~\mathrm{(stat)})\%$ for data and Monte Carlo, respectively. The two values are consistent within the uncertainties.

The reconstruction efficiency presented here only assesses whether a track is found and reconstructed. The efficiency here does not quantify the overall quality of the track reconstruction in the TPC, such as the correctness of the track length or angle; this will be addressed in a future publication.

\begin{figure}[htbp]
  \begin{center}
    \includegraphics[width=0.7\linewidth]{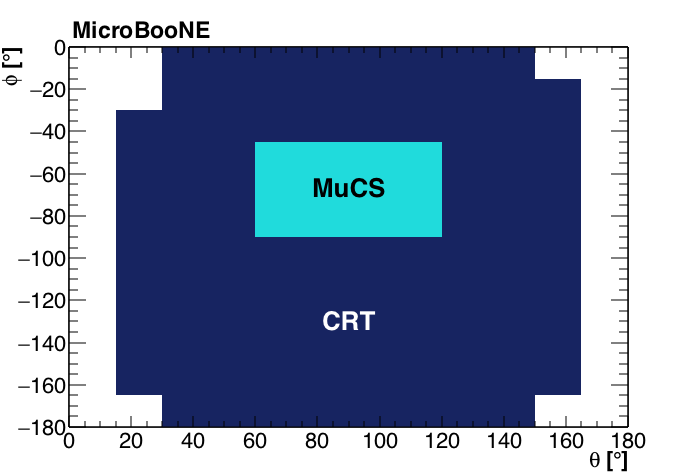}
    \caption{Monte Carlo simulation of the coverage in the ($\theta,\phi$) plane for both the MuCS (light blue), as presented in this article, and the CRT system (dark blue).} \label{fig:crt}
  \end{center}
\end{figure}

We also analyze systematic uncertainties that affect the data reconstruction efficiency. The fraction of muons triggering the MuCS that decay or are captured before reaching the TPC is~$\approx1.0\%$, according to a Monte Carlo simulation.  This factor is taken into account in the measurement of the data reconstruction efficiency.
The MuCS data are acquired with the two MuCS boxes placed in three different positions, covering different parts of the detector. Detector non-uniformities induce a non-negligible effect ($\sim$1\%) in the measurement of the data reconstruction efficiency. Other systematic uncertainties, such as the ones related to the space-charge effect and the energy sampling, were found to be negligible with the present level of statistics.
This article describes a proof of principle method of using an external muon counter to measure the cosmic-ray reconstruction efficiency in a LArTPC. The ($\theta, \phi, L$) parameter space covered by the MuCS will be significantly expanded using the data coming from a larger cosmic ray tagger system (CRT), installed in March 2017, as illustrated in figure~\ref{fig:crt}. This detector will be able to tag $\approx$80\% of the cosmic rays hitting the TPC and study the presence of non-uniformities in a larger portion of the MicroBooNE detector. The data coming from the CRT will allow the measurement of efficiency-corrected quantities, such as the cosmic-ray flux, and the reconstruction efficiencies will be directly applicable to physics measurements.

\clearpage{}

\acknowledgments

This material is based upon work supported by the following: the U.S. Department of Energy, Office of Science, Offices of High Energy Physics and Nuclear Physics; the U.S. National Science Foundation; the Swiss National Science Foundation; the Science and Technology Facilities Council of the United Kingdom; and The Royal Society (United Kingdom). Fermilab is operated by Fermi Research Alliance, LLC under Contract No. DE-AC02-07CH11359 with the United States Department of Energy. The Muon Counter Stack and its dedicated DAQ were provided by the Virginia Polytechnic Institute and State University externally to the MicroBooNE collaboration using spare electronics from the Double Chooz experiment provided by Columbia University's Nevis Laboratories.


\end{document}